\documentstyle[12pt,graphicx,epsf]{article}
\begin{document}

\def\NPB#1#2#3{{\it Nucl.\ Phys.}\/ {\bf B#1} (#2) #3}
\def\PLB#1#2#3{{\it Phys.\ Lett.}\/ {\bf B#1} (#2) #3}
\def\PRD#1#2#3{{\it Phys.\ Rev.}\/ {\bf D#1} (#2) #3}
\def\PRL#1#2#3{{\it Phys.\ Rev.\ Lett.}\/ {\bf #1} (#2) #3}
\def\PRT#1#2#3{{\it Phys.\ Rep.}\/ {\bf#1} (#2) #3}
\def\MODA#1#2#3{{\it Mod.\ Phys.\ Lett.}\/ {\bf A#1} (#2) #3}
\def\IJMP#1#2#3{{\it Int.\ J.\ Mod.\ Phys.}\/ {\bf A#1} (#2) #3}
\def\nuvc#1#2#3{{\it Nuovo Cimento}\/ {\bf #1A} (#2) #3}
\def\RPP#1#2#3{{\it Rept.\ Prog.\ Phys.}\/ {\bf #1} (#2) #3}
\def\APJ#1#2#3{{\it Astrophys.\ J.}\/ {\bf #1} (#2) #3}
\def\APP#1#2#3{{\it Astropart.\ Phys.}\/ {\bf #1} (#2) #3}
\def\etal{{\it et al\/}}

\newcommand{\bev}{\begin{verbatim}}
\newcommand{\beq}{\begin{equation}}
\newcommand{\beqa}{\begin{eqnarray}}
\newcommand{\beqn}{\begin{eqnarray}}
\newcommand{\eeqn}{\end{eqnarray}}
\newcommand{\eeqa}{\end{eqnarray}}
\newcommand{\eeq}{\end{equation}}
\newcommand{\Eev}{\end{verbatim}}
\newcommand{\bec}{\begin{center}}
\newcommand{\eec}{\end{center}}
\def\ie{{\it i.e.}}
\def\eg{{\it e.g.}}
\def\half{{\textstyle{1\over 2}}}
\def\nicefrac#1#2{\hbox{${#1\over #2}$}}
\def\third{{\textstyle {1\over3}}}
\def\quarter{{\textstyle {1\over4}}}
\def\m{{\tt -}}
\def\mass{M_{l^+ l^-}}
\def\p{{\tt +}}

\def\slash#1{#1\hskip-6pt/\hskip6pt}
\def\slk{\slash{k}}
\def\GeV{\,{\rm GeV}}
\def\TeV{\,{\rm TeV}}
\def\y{\,{\rm y}}

\def\l{\langle}
\def\r{\rangle}

\begin{titlepage}
\samepage{
\setcounter{page}{1}
\rightline{UNILE-CBR-2001-3}
\rightline{OUTP-01-33P}
\vspace{1.5cm}
\begin{center}
 {\Large \bf Susy QCD and High Energy Cosmic Rays 1.\\}
\vspace{.5 cm}
{\Large \bf  Fragmentation functions of Susy QCD \\  }
\vspace{1.cm}
 {\large Claudio Corian\`{o}\footnote{ E-mail address: Claudio.Coriano@le.infn.it}
 and Alon E. Faraggi\footnote{faraggi@thphys.ox.ac.uk}\\
\vspace{.5cm}
{\it $^1$Dipartimento di Fisica\\
 Universita' di Lecce\\
 I.N.F.N. Sezione di Lecce \\
Via Arnesano, 73100 Lecce, Italy\\}
\vspace{.5cm}
{\it $^2$Theoretical Physics Department\\
University of Oxford, Oxford, OX1 3NP, United Kingdom}}
\end{center}
\begin{abstract}
The supersymmetric evolution of the fragmentation functions (or timelike evolution) 
within $N=1$ $QCD$ is discussed and predictions for the fragmentation functions 
of the theory (into final protons) are given. 
We use a backward running of the supersymmetric DGLAP equations, 
using a method developed in previous works. 
We start from the usual QCD parameterizations at low energy and run the DGLAP back, 
up to an intermediate scale -assumed to be supersymmetric- where we switch-on supersymmetry. 
From there on we assume the applicability of an $N=1$ supersymmetric evolution (ESAP). 
We elaborate on possible application of these results to High Energy 
Cosmic Rays near the GZK cutoff.

\end{abstract}
\smallskip}
\end{titlepage}

\section{Introduction}

The Standard Model of elementary particle physics successfully accounts for
all the observed particle properties. Its remarkable success 
suggests the possibility that it correctly describes the particle 
properties up to energy scales which far exceed the currently accessible energy scales.
Indeed, this possibility has been entertained in the explorations of 
Grand Unified, and Superstring, Theories. Additionally, the spectacular
confirmation of the Standard Model in high energy colliders, as well
as the proton longevity and suppression of left--handed neutrino masses,
provide strong support for the grand desert scenario and unification. 
The question then is how are we going to find out whether this is 
indeed the path chosen by nature, as direct signatures at LHC/NLC 
will be able to probe the desert only up to a few TeV.
Therefore we need new tools to achieve this. 
An example of such a probe that has been used with great success is that
of the gauge coupling unification. The general methodology in this respect
is to extract the gauge couplings from experiments that are performed
at the accessible energy scales. Based on concrete assumptions in 
regard to the particle content in the desert, the couplings are then 
extrapolated to the high energy scale and the unification hypothesis is 
tested. In this manner the consistency of the specific assumptions in 
regard to the particle content in the desert with the hypothesis of the 
gauge coupling unification is subjected to an experimental test. 
The well known spectacular success of this methodology is in
differentiating between gauge coupling unification in supersymmetric 
and non--supersymmetric Grand Unified Theories
\cite{gcurge}. The unification hypothesis is
found not to be consistent with the low energy data, unless the
minimal particle content is supersymmetric, or is modified
in some other way. Our task then is to develop additional
tools that can serve as similar useful probes of the desert. 

\subsection{ Motivations for this work}
In this paper we start applying this philosophy to another physical setting.

A different experimental observation - which also points toward 
the validity of the big desert scenario - is the existence of Ultra High Energy
Cosmic Rays above the Greisen--Zatsepin--Kuzmin cutoff. 
A plausible explanation for the observation of such cosmic rays
is the existence of supermassive metastable states with mass
of the order $10^{12}-10^{13}{\rm GeV}$, and a lifetime that exceeds
$10^{10}y$. The possibility that
such states compose a substantial component of the dark matter
as well as that they may explain the observation of UHECR 
has been discussed elsewhere. It has been further shown
that realistic string models often give rise to states 
with precisely such properties \cite{ccf}.

The existence of such super--massive, slowly decaying states,
may therefore serve as a probe of the grand desert.
A plausible explanation of the observed showers in the UHECR
is that the original supermassive decaying particle decays 
into strongly interacting particles which subsequently fragment 
into hadrons \cite{Subir1}. This hypothesis therefore needs 
the relevant fragmentation functions at the energy scale
of the originally decaying particle. These functions,
however, similarly to the case of the gauge couplings considered in much of the 
literature on unification, are extracted from 
experiments at the currently accessible (low) energy scales which are much 
below that of the decaying particle. 
It is obvious that this points toward the use of the renormalization group (RG) 
evolution. 
However, this
evolution, as well as the composition of the produced shower,
depends on the assumptions made in regard to the composition of
the spectrum in the desert. We conclude that similarly to the gauge coupling
unification methodology, the evolution of the fragmentation functions
by utilizing the renormalization group extrapolation, and the
subsequent implementation in the analysis of the UHECR hadronic
showers, can serve as a tool that differentiates between different
assumptions on the very high energy spectrum.

With these motivations in mind, here we develop the instruments to evolve the fragmentation functions
in supersymmetric QCD up to a high energy scale.
In particular, in this paper we present 
for the first time the structure of the supersymmetric DGLAP 
(Exact Supersymmetric DGLAP or ESAP) 
equations in the timelike region. 
We solve them numerically for all flavours
and introduce various combinations of non-singlet matrix equations 
to reach this result. We then study the impact of the 
new evolution assuming a common scale for susy breaking and restoration, 
parameterized by the mass of the superpartners, here assumed to be mass degenerate.

\subsection{Step Approximations} 
Given the long stretch it takes to proceed with the analysis 
of the susy evolution, a topic of numerical complexity in its own, 
the applications to cosmic rays of our work will be presented in a companion 
paper that will follow shortly. We also have decided to address issues related to 
the fragmentation region of $N=1$ QCD starting from the ``low energy'' end, 
since nothing is known of these functions at any scale. The term ``low energy'' 
(the results we discuss here are obtained in the $10^{3}- 10^{8}$ GeV energy range )
- from the point of view of collider phenomenology - is a misnomer, 
but it is not in the context of high energy cosmic ray physics.  
There is also another more direct reason for limiting our analysis to this range.
We have discovered some features of the RG evolution which require 
a special care at such large energies. We find -in our numerical studies- 
the onset of an instability in the evolution in the singlet sector of the fragmentation which requires an independent investigation. It is not clear to us, at this point, 
whether this instability is an intrinsic limitation of the algorithm used 
in the numerical implementation or if the perturbative 
expansion needs a resummation. At such large scales this last option remains open. 
We should also mention that we work in ``x'' space, and this might be a limitation. 
We hope to return to this issue in the near future.

 The analysis of fragmentation functions of Susy QCD is a topic 
of remarkable phenomenological interest both 
for collider phenomenology and in the astroparticle context. 
While - recently - a detailed analysis of the 
supersymmetric evolution of the parton distributions of Susy QCD has been presented 
\cite{CC1,CC2,CC3}, the study of the evolution of the supersymmetric 
fragmentation functions is still missing. Of particular interest 
is the study of a combined QCD/SQCD evolution with intermediate 
regions in the evolution 
characterized by partial supersymmetry (SAP) or exact supersymmetry (ESAP) \cite{CC2}. 
It is well known that a matching 
between these regions is possible with specific boundary conditions. These 
boundary conditions should come from some 
deeper understanding of the way in which supersymmetry is broken and restored 
as we run into the different stages of the renormalization group evolution. 
In a first approximation, however, it is possible to assume 
that the regular QCD distributions (or fragmentation) functions are continuous 
at each intermediate region, 
thereby neglecting threshold effects which can't be inferred 
from first principles. As we cross any supersymmetric region, starting, 
for instance, from the end of the regular QCD evolution, 
it is possible to assume that supersymmetric distributions and fragmentation functions are generated radiatively. We recall that a similar modeling  
-which neglects possible effects of higher twist at the thresholds- is well spread in the 
case of ordinary QCD.
\section{Initial/Final State Scaling Violations}
\subsection{The Spacelike Evolution}
The first analysis of the spacelike evolution was
carried out in ref.\cite{KR}, 
where simple models -obtained from the analysis of the first 2 moments - 
of the supersymmetric parton distributions were also presented. 
A complete strategy to solve these equations and the strategy to generate 
supersymmetric scaling violations in N=1 QCD was put forward in Refs.~\cite{CC1,CC2}.  
In $N=1$ QCD gluons have partners called gluinos (here denoted by $\lambda$)
and left- and right-handed quarks have complex scalar partners (squarks) which we denote as 
$ \tilde{q_L}$ and $\tilde{q_R}$ with $\tilde{q}=\tilde{q}_L + \tilde{q}_R$ (for left-handed and right-handed squarks respectively). 

The interaction between the elementary fields are described by the 
$SU(3)$ color gauge invariant and supersymmetric Lagrangian 

\beqa
&&{ \cal L}= -{1\over 4} G_{\mu\nu}^a G^{\mu\nu}_a + 
{1\over 2}\bar{\lambda}_a \left( i \not{D}\right) \lambda_a \nonumber \\
&& + i \bar{q}_i \not{D} q_i + D_\mu \tilde{q}_R D^\mu\tilde{q}^{R}
+ D_\mu \tilde{q}_L D^\mu\tilde{q}_{L} + i g \sqrt{2}
\left(\bar{\lambda^a}_R \tilde{q}_{i L}^\dagger  T^a q_{L i} + 
\bar{\lambda^a}_L \tilde{q}_{i R}^\dagger  T^a q_{R i}  - {\tt h.\,\,\,c.} \right)
\nonumber \\
&& -{1\over 2} g^2 \left( \tilde{q}_{L i}^\dagger T^a \tilde{q}_{L i}
- \tilde{q}_{R i}^\dagger T^a \tilde{q}_{R i}^\dagger\right)^2 
+ {\tt mass\,\,\,\, terms },\nonumber \\
\eeqa

where $a$ runs over the adjoint of the color group and $i$ denotes the flavours.

$\tilde{q}_{iR}$ are the supersymmetric partners of right-handed quarks $q_{iR}$ and 
$\tilde{q}_{iL}$ are those of the left-handed quarks $q_{iL}$. 

We remind here that in actual phenomenological applications, 
it is useful to extract the light-cone dynamics -or parton model picture- 
of supersymmetric collisions by invoking a factorization of the cross section 
into hard and soft contributions. Fragmentation functions appear in this context.  
An integral part of this analysis is the use of renormalization group equations -susy DGLAP- 
which resum the logarithmic scaling violations and evolve distributions functions 
and fragmentation functions to the appropriate factorization scale.

 These studies require the matrix of the anomalous dimensions (for all the moments), or of the corresponding Altarelli-Parisi (DGLAP) kernels. 
We recall that the leading order anomalous dimensions are known \cite{Antoniadis,KR} both 
for a partial susy evolution and for the exact one. 
\subsection{The Timelike Evolution}
In this work we focus our attention on the evolution of 
fragmentation functions in the context of exact supersymmetry, 
with coupled gluons and squarks. 
\begin{figure}
\centerline{\includegraphics[angle=0,width=.7\textwidth]{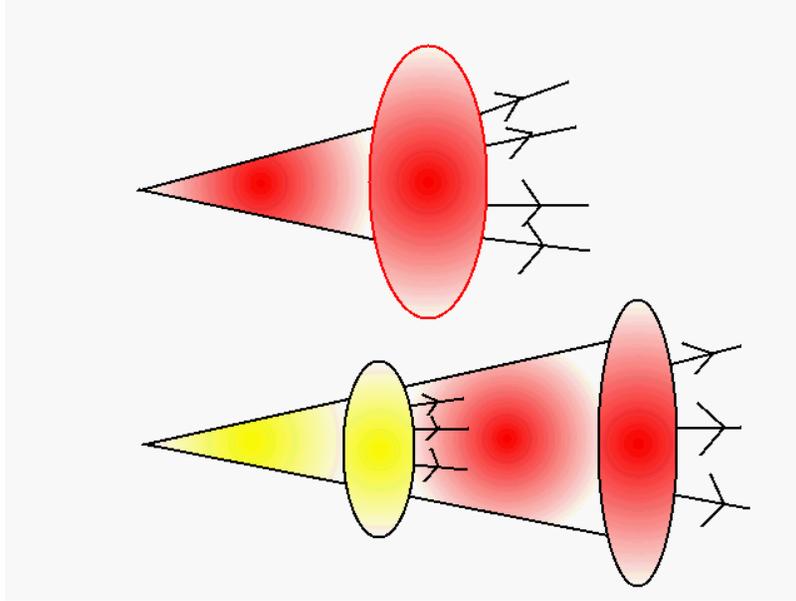}}
\caption{An illustration of the regular versus a mixed supersymmetric evolution}
\end{figure}
Fig.~1 summarizes the basic strategy of our work. The figure at the top 
depicts a regular evolution (a one-phase evolution), while the one at the bottom 
illustrates a mixed QCD/SQCD evolution (a two-phase evolution). The first stage 
is supersymmetric, the second one is regular. In the first stage all the partons 
(susy and not susy) are massless, in the second one only quark and gluons survive. 
At the end of the first stage the supersymmetric partners are extinct, 
but their presence at the higher scales is saved in the boundary conditions for the 
quarks and gluons fragmentation functions, when the second stage of the evolution starts. 
 
Within the approximation implied by the use of a pure SQCD evolution, no missing 
$E_t$ events are allowed along the development of the supersymmetric cascade, 
since electroweak and flavour mixing effects are not included. Their inclusion is still 
an open chapter even in (QCD) Monte Carlo event generators for the final state.

The ladder structure of the supersymmetric showers in the leading logarithmic 
approximation can be easily pictured in terms of symmetric doubling of lines 
of regular QCD ladders in all the possible allowed cases. 
Showers can be equally initiated by $q\,\, \bar{q}$ pairs, gluon pairs, gluino pairs or 
squark pairs. The supersymmetric transition (i.e. those involving supersymmetric 
partners) stop (see Fig.~1 (bottom)) once the supersymmetry breaking scale 
$M_{\lambda}=M_{\tilde{q}}$ is reached.

Similarly to the QCD case, in the case of exact 
$N=1$ supersymmetry we define singlet and non-singlet fragmentation functions $D^h_f(x,Q^2)$. 
They describe the amplitude for a parton of type $f$
to fragment into a hadron $h$ as a function 
of the Bjorken variable $x$ (fractional energy of the fragment) 
and initial energy $Q$. Their operatorial definition is similar 
to that of ordinary parton distributions. The evolution equations for these functions 
are similar to the standard DGLAP equations, but with transposed kernel matrix 
$(P\to P^T)$. In leading order the analytic continuation (from the spacelike 
or DIS evolution to the timelike evolution) is straightforward, while 
some complications appear to next-to-leading order. They involve a breaking of the Drell-Yan-Levy relation, which appears to be 
violated at parton level \cite{blumlein}. Since our analysis is in the leading logarithmic approximation, 
we will not consider these aspects in our work any further.

The equations for the timelike evolution are given by 

\beqa
\frac{d}{d\log(Q^2)}D^h_g(x,Q^2) &=&
\frac{\alpha_s}{2\pi}\left(P_{gg}\otimes D^h_g + P_{\lambda g}\otimes D^h_\lambda + 
P_{qg}\otimes \sum_i\left(D^h_{q_i} + D^h_{\bar{q}_i}\right)\right. \nonumber \\
&&\left.\,\,\,\,\,\,\,\,\,\,\,\, + P_{\tilde{q} g}\otimes\sum_{i=1}^{n_f}\left(\tilde{q}_{iL} + \tilde{q}_{iR} 
+ \bar{\tilde{q}}_{iL} + \bar{\tilde{q}}_{iR}\right)\right) \\
\frac{d}{d\log(Q^2)}D^h_\lambda(x,Q^2) &=&
\frac{\alpha_s}{2\pi}\left(P_{g\lambda}\otimes D^h_g + P_{\lambda \lambda}\otimes D^h_\lambda + 
P_{q \lambda}\otimes \sum_i\left({D^h_{q_i} + D^h_{\bar{q}_i}}\right)\right.\nonumber \\
 &&\left.\,\,\,\,\,\,\,\,\,\,\,\, + P_{\tilde{q}\lambda}\otimes \sum_{i=1}^{n_f}\left( \tilde{q}_{iL} + \tilde{q}_{iR} + \bar{\tilde{q}}_{iL} 
+ \bar{\tilde{q}}_{iR}\right)\right)\\
\frac{d}{d\log(Q^2)}D^h_{q_i}(x,Q^2) &=&
\frac{\alpha_s}{2\pi}\left(\frac{1}{2 n_f}P_{g q}\otimes D^h_g + \frac{1}{2 n_f}P_{\lambda q}\otimes D^h_\lambda + 
P_{\tilde{q} q }\otimes \left( D^h_{{\tilde{q}}_{iL}} + 
D^h_{{\tilde{q}}_{iR}}  \right)\right.\nonumber \\
 &&\left. \,\,\,\,\,\,\,\,\,\,\,\, + P_{qq}\otimes D^h_{{q_i}}\right)  \\
\frac{d}{d\log(Q^2)}D^h_{{\tilde{q}}_{iL}} (x,Q^2) &=&
\frac{\alpha_s}{2\pi}\left(\frac{1}{4 n_f}P_{g\tilde{q} }\otimes D^h_g + \frac{1}{4 n_f}P_{\lambda \tilde{q}}\otimes 
D^h_\lambda + \frac{1}{2}P_{q\tilde{q} }\otimes D^h_{{q}_{i}}\right.\nonumber \\
 &&\left.\,\,\,\,\,\,\,\,\,\,\,\, + P_{\tilde{q}\tilde{q}}\otimes D^h_{\tilde{q}_{iL}}\right) \\
\frac{d}{d\log(Q^2)}D^h_{\tilde{q}_{iR}}(x,Q^2) &=&
\frac{\alpha_s}{2\pi}\left(\frac{1}{4 n_f}P_{g\tilde{q} }\otimes D^h_g + \frac{1}{4 n_f}
P_{\lambda \tilde{q}}\otimes 
D^h_\lambda +\frac{1}{2}P_{q\tilde{q} }\otimes D^h_{q_i}\right.\nonumber \\
 &&\left.\,\,\,\,\,\,\,\,\,\,\,\, + P_{\tilde{q}\tilde{q}}\otimes D^h_{\tilde{q}_{iR}}\right)\nonumber \\
\eeqa
In the following we will use the short-hand notation 
\beq
D^h_{\tilde{q}_i}(x,Q^2)\equiv D^h_{\tilde{q}_{iL}}(x,Q^2) + D^h_{\tilde{q}_{iR}}(x,Q^2)
\eeq
to denote the fragmentation functions of squarks of flavour $i$ 
at a fractional energy $x$ and momentum $Q$. It is also convenient to separate 
the equations, as usual, into singlet and non singlet sectors using the definitions

\beqa
 D^h_{q_V}(x, Q^2) &\equiv& \sum_{i=1}^{n_f}\left( D^h_{{q}_i}(x,Q^2) - D^h_{{\bar{q}}_i}(x, Q^2)\right) , \nonumber \\
&\equiv & D^h_{q_{NS}}\nonumber \\ 
 D^h_{\tilde{q}_V}(x, Q^2) &=& \sum_{i=1}^{n_f}
\left( D^h_{\tilde{q}_i}(x, Q^2) - D^h_{\tilde{\bar{q}}_i}(x, Q^2)\right) \nonumber \\
& \equiv &D^h_{\tilde{q}_{NS}}(x,Q^2)\nonumber \\ 
 D_{q^{+}}(x, Q^2) &\equiv& \sum_{i=1}^{n_f} \left(D^h_{{q}_i}(x, Q^2) + D^h_{\bar{q}_i}(x, Q^2)\right) \nonumber \\
D^h_{\tilde{q}^{+}}(x, Q^2)&\equiv& \sum_{i=1}^{n_f}\left( D^h_{\tilde{q}_i}(x,Q^2) + D^h_{\tilde{\bar{q}}_i}(x,Q^2)\right).
\eeqa

The non singlet equations are 
\beqa
Q^2 {d\over d Q^2} D{q_V}(x,Q^2)&=&{\alpha(Q^2)\over 2 \pi} \left( P_{qq}\otimes D^h_{q_V} + P_{\tilde{q}q}\otimes 
D_{\tilde{q_V}}\right)  \nonumber \\
Q^2 {d\over d Q^2} D_{\tilde{q}_V}(x,Q^2)&=&{\alpha(Q^2)\over 2 \pi} \left( P_{q\tilde{q}}\otimes D^h_{q_V} +P_{\tilde{q}\tilde{q}}\otimes {D^h_{\tilde{q_V}}}\right),  \nonumber \\
\label{susyns}
\eeqa

and the singlet matrix equations, which mix $q_V$ and $\tilde{q}_V$ with the gluons and the gluinos 

\beq
 Q^2 {d\over d Q^2}  \left[\begin{array}{c}   D^h_g(x,Q^2)\\ 
D^h_{\lambda}(x,Q^2) \\ D^h_{q^+}(x,Q^2) \\D^h_{\tilde{q}^+}(x,Q^2)  \end{array} \right]
=\left[ \begin{array}{llll} 
         P_{gg} & P_{g \lambda} & P_{g q} &  P_{g \tilde{q}} \\
         P_{\lambda g} & P_{\lambda \lambda} & P_{\lambda q} & P_{\lambda \tilde{q}} \\
         P_{q g} & P_{q \lambda} & P_{q q} & P_{q \tilde{q}} \\
         P_{\tilde{q} g} & P_{\tilde{q} \lambda} & P_{\tilde{q} q} & P_{\tilde{q} \tilde{q}} 
         \end{array} \right]^T \otimes
  \left[ \begin{array}{c}        
         D^h_g(x,Q^2)\\ D^h_\lambda(x,Q^2) \\ D^h_{q^{(+)}}(x,Q^2) \\ D^h_{\tilde{q}^{(+)}}(x,Q^2) \end{array} \right].
\eeq
where $``T''$ indicates the matrix transposed.  
To solve for all the flavours, it is convenient to introduce the linear combinations 
\beqa
\chi_i(x,Q^2) &=& D^h_{q_i^{(+)}} -\frac{1}{n_f}D^h_{q^{(+)}} \nonumber \\
\tilde{\chi}_i(x,Q^2) &=& D^h_{\tilde{q}_i^{(+)}} -\frac{1}{n_f} 
D^h_{\tilde{q}^{(+)}} \nonumber \\
\eeqa
and the additional singlet equations 

\beqa
Q^2 {d\over d Q^2} D_{q_i^{(-)}}(x,Q^2)&=&
{\alpha(Q^2)\over 2 \pi} \left( P_{qq}\otimes D^h_{q_i^{(-)}} + P_{\tilde{q}q}\otimes 
D_{\tilde{q_i}^{(-)}}\right)  \nonumber \\
Q^2 {d\over d Q^2} D_{\tilde{q}_i^{(-)}}(x,Q^2)&=&
{\alpha(Q^2)\over 2 \pi} \left( P_{q\tilde{q}}\otimes D^h_{q_i^{(-)}} 
+P_{\tilde{q}\tilde{q}}\otimes {D^h_{\tilde{q}_i^{(-)}}}\right)  \nonumber \\
\label{susyns}
\eeqa
and 
\beqa
Q^2 {d\over d Q^2} D^h_{\chi_i}(x,Q^2)&=&
{\alpha(Q^2)\over 2 \pi} \left( P_{qq}\otimes D^h_{\chi_i} + P_{\tilde{q}q}\otimes 
D^h_{\tilde{\chi_i}}\right)  \nonumber \\
Q^2 {d\over d Q^2} D^h_{\tilde{\chi}_i}(x,Q^2)&=&
{\alpha(Q^2)\over 2 \pi} \left( P_{q\tilde{q}}\otimes D^h_{\chi_i} 
+P_{\tilde{q}\tilde{q}}\otimes D^h_{\tilde{\chi_i}}\right).  \nonumber \\
\label{susyns}
\eeqa
The general flavour decomposition is obtained by solving the singlet equations for 
$D^h_{q^{(+)}}$ and $D^h_{\tilde{q}^{(+)}}$, then solving the non-singlet equations 
for $D^h_{q_i^{(-)}}$ and $D^h_{\tilde{q}_i^{(+)}}$ and for $D^h_{\chi_i}$ 
and $D^h_{\tilde{\chi}_i}$. The fragmentation functions of the various flavours are extracted using the linear combinations 

\beqa
D^h_{q_i} &=& \frac{1}{2}\left(D^h_{q_i^{(-)}} + D^h_{\chi_i} +\frac{1}{n_f}D^h_{q^{(+)}}\right) 
\nonumber \\
D^h_{\bar{q}_i} &=& -\frac{1}{2}\left(D^h_{q_i^{(-)}} - D^h_{\chi_i} -\frac{1}{n_f}D^h_{q^{(+)}}\right) \nonumber \\
\eeqa

and 
\beqa
D^h_{\tilde{q}_i} &=& \frac{1}{2}\left(D^h_{\tilde{q}_i^{(-)}} + D^h_{\tilde{\chi}_i} 
+\frac{1}{n_f}D^h_{\tilde{q}^{(+)}}\right) 
\nonumber \\
D^h_{\bar{{\tilde{q}}}_i} &=& -\frac{1}{2}\left(D^h_{\tilde{q}_i^{(-)}} 
- D^h_{\tilde{\chi}_i} -\frac{1}{n_f}D^h_{\tilde{q}^{(+)}}\right) \nonumber \\
\eeqa
to identify the various flavour components.

There are simple ways to calculate the kernel of the susy DGLAP evolution by a simple extension 
of the usual methods. 
The changes are primarily due to color factors. There are also some
basic supersymmetric relations which have to be satisfied \cite{CC1}. 
They are generally broken in the case of decoupling.  
We recall that the supersymmetric version of the $\beta$ function is given at two-loop level by

\beqa
\beta_0^S &=& \frac{1}{3}\left(11\, C_A - 2\, n_f - 2 \,n_\lambda \right) 
\nonumber \\
\beta_1^S &=&\frac{1}{3}\left( 34\, C_A^2 - 10\, C_A\, n_f - 10\, C_A\, n_\lambda 
- 6\, C_F\, n_f - 6 \,C_\lambda\, n_\lambda \right)
\eeqa
where $n_f$ is the number of flavours and 
for Majorana gluinos $n_\lambda=C_A$.
 
The ordinary running of the coupling is replaced by its supersymmetric running

\beq
{\alpha^S(Q_0^2)\over 2 \pi}={2\over \beta_0^S}
{ 1\over \ln (Q^2/\Lambda^2)}\left( 1 - {\beta_1^S\over \beta_0^S}
{\ln\ln(Q^2/\Lambda^2)\over \ln(Q^2/ \Lambda^2)} +
O({1\over\ln^2(Q^2/\Lambda^2) })\right).
\eeq

The kernels are modified both in their coupling $(\alpha \to \alpha^S)$ 
and in their internal structure (Casimirs, color factors, etc.) when moving 
from the QCD case to the SQCD case. In order 
to illustrate the approach that we are going to follow in our analysis of the 
fragmentation functions within a mixed SQCD/QCD evolution, 
we recall the strategy employed in Refs.~\cite{CC1,CC2} to generate the ordinary 
supersymmetric distribution functions. 

In Ref.~\cite{CC2} for scaling violations affecting the initial state were introduced 3 
regions: 
\begin{itemize}
\item 1) the QCD region, described by ordinary QCD (Altarelli Parisi or AP) ;

\item 2) an intermediate supersymmetric region with coupled gluinos 
and decoupled squarks (partially supersymmetric AP or SAP); 

\item 
3) the N=1 region (exact supersymmetric AP or ESAP). 
\end{itemize}
The basic idea was to generate these distributions {\em radiatively} as usually 
done in QCD for the gluons, for instance, using the fact that the matrix of the anomalous 
dimensions is not-diagonal. 
The most general sequence of evolutions   
(denoted AP-SAP-ESAP in \cite{CC2}) is described by the arrays 
$(Q_i,Q_f)_{AP}-(Q_i,Q_f)_{SAP}-(Q_i,Q_f)_{ESAP}$, with $Q_{f,AP}=Q_{i,SAP}=m_{2\lambda}$ 
and $Q_{f,SAP}=Q_{i,ESAP}=m_{\tilde{q}}$. In this work we limit our analysis to a 
simpler AP-ESAP evolution and we run the DGLAP equations {\em back} starting 
from known QCD fragmentation functions - for which 
various sets are available in the literature \cite{kkp} - 
up to an intermediate supersymmetric scale. From this point on (up in energy) we switch 
on the ESAP evolution of fragmentation functions. As we run the equations upward, 
supersymmetric fragmentation functions are generated. The boundary values of the 
low energy (QCD) functions at the supersymmetry scale $m_{2\lambda}$ set 
the initial condition  for the (backward) supersymmetric running 
up to the final scale. In simple terms: we reach the mountain 
from the valley, and cross a fence along the way. 

In general, if we split the intermediate susy scale into 2 sectors 
$m_{\tilde{q}} << m_{2\lambda}$ (ESAP-SAP-AP evolution) 
in this (general) case the solution is built by sewing the three regions as  

\beqa
\left[ \begin{array}{c} 
         D^h_{q_V}(x,Q^2)\\
         D^h_{\tilde{q}_V}(x,Q^2)        \\	
         \end{array} \right] &=& \left[ \begin{array}{c} 
         D^h_{q_V}(x,Q_0^2)\\
         0        \\	
         \end{array} \right] + \int_{Q_0^2}^{m_{2\lambda}^2} d\, \log\, Q^2\, 
P_{AP}^{T,NS}(x,\alpha(Q^2))\otimes \left[ \begin{array}{c} 
         D^h_{q_V}(x,Q^2)\\
         0        \\	
         \end{array} \right] \nonumber \\
&+& \int_{m_{2\lambda}^2}^{m_{ \tilde{q}}^2} d\, \log Q^2 
P_{SAP}^{T,NS}(x,\alpha^S(Q^2))\otimes \left[ \begin{array}{c} 
         D^h_{q_V}(x,Q^2)\\
         0        \\	
         \end{array} \right] \nonumber \\
&+& \int_{m_{2\lambda}^2}^{Q_f^2} d\, \log Q^2\, 
P_{ESAP}^{T,NS}(x,\alpha^{ES}(Q^2))\otimes \left[ \begin{array}{c} 
         D^h_{q_V}(x,Q^2)\\
         D^h_{\tilde{q}^V}(x,Q^2)        \\	
         \end{array} \right] \nonumber \\
\eeqa
in the non singlet and

\beqa
\left[ \begin{array}{c} 
         D^h_{G}(x,Q_f^2)\\
	 D^h_{\lambda}(x,Q^2)\\
	 D^h_{q^+}(x,Q^2)\\
         D^h_{\tilde{q}^+}(x,Q^2)\\	
         \end{array} \right] &=& \left[ \begin{array}{c} 
         D^h_G(x,Q_f^2)\\
	 0\\
	  D^h_{q^+}(x,Q_f^2) \\
         0\\	
         \end{array} \right]  + \int_{Q_0^2}^{m_{2\lambda}^2} d\, \log\, Q^2\, 
P_{T,AP}^{NS}(x,\alpha(Q^2))\otimes 
\left[ \begin{array}{c} 
         D^h_G(x,Q_f^2)\\
	  0\\
	 D^h_{q^+}(x,Q^2)\\
         0\\	
         \end{array} \right] \nonumber \\
&+& \int_{m_{2\lambda}^2}^{m_{ \tilde{q}}^2} d\, \log Q^2 
P_{T,SAP}^{NS}(x,\alpha^S(Q^2))\otimes 
\left[ \begin{array}{c} 
         D^h_G(x,Q_f^2)\\
	 D^h_{\lambda}(x,Q^2)\\
	 D^h_{q^+}(x,Q^2)\\
           0\\	
         \end{array} \right] \nonumber \\
&+& \int_{m_{2\lambda}^2}^{Q_f^2} d\, \log Q^2\, 
P_{T,ESAP}^{NS}(x,\alpha^{ES}(Q^2))\otimes
\left[ \begin{array}{c} 
         D^h_G(x,Q_f^2)\\
	 D^h_{\lambda}(x,Q^2)\\
	 D^h_{q^+}(x,Q^2)\\
           D^h_{\tilde{q}^+}(x,Q^2)\\
         \end{array} \right] \nonumber \\
\eeqa
for the singlet solution. 
The zero entries in the arrays for some of the distributions are due 
to the boundary conditions, since all the supersymmetric partners are generated, in this model, 
by the evolution. 

The general structure of the algorithms that solves these equations 
is summarized below. We start from the non singlet sector
and then proceed to the singlet. 
{}From now on, to simplify our notation, we will omit at times the
index ``$h$'' from the fragmentation functions when obvious. 
We define 
$D_{q^{NS}}(x,Q^2)=\left(D_{q_V}(x,Q^2),D_{\tilde{q}_V}(x,Q^2)\right)^T$  and 
$DA_n(x)=\left(DA^{q_V}_n,DA^{\tilde{q}_V}_n\right)^T$ and introduce the ansatz

\beq
D_{q^{NS}}(x,Q^2)=\sum_{n=0}^{n_0}\,{A_n(x)\over n!} \log^n
\left(\frac{\alpha(Q^2)}{\alpha{(Q_0^2)}}\right),
\eeq
where $n_0$ is an integer at which we stop the iteration, which 
usually ranges up to 20.
The first coefficient of the recursion is determined by the initial condition
\beq
DA_0(x) = U(x)\otimes D_q^{NS}(x,Q_0^2), 
\eeq
where
\beqa
U_1(x) &\equiv & 
\left[ \begin{array}{ll} 
         \delta(1-x) & 0\\
	  0 & 0\\	
         \end{array} \right] \nonumber \\
U_{1\,2}(x) &\equiv & 
\left[ \begin{array}{ll} 
         \delta(1-x) & 0\\
	  0 & \delta(1-x)\\	
         \end{array} \right]. 
\eeqa

The recursion relations are given by 

\beq
DA_{n+1}(x)= -\frac{2}{\beta_0}P^{T,NS}_{AP}\otimes DA_n(x)
\eeq
The solution in the first (DGLAP) region at the first matching scale 
$m_{2\lambda}$ is given by 
\beq
D_{q^{NS}}(x,m_{2\lambda})=\sum_{n=0}^{n_0}\frac{DA^{AP}_n(x)}{n!}\log^n
\left(\frac{\alpha({m_{2\lambda}}^2)}{\alpha(Q_0^2)}\right).
\eeq
At the second stage the (partial) supersymmetric coefficients are given by 
(S is a short form of $SAP$)

\beqa
DA^{S,NS}_0(x) &=& U(x)\otimes D_q(x,m_{2\lambda^2}) \nonumber \\
DA^{S,NS}_{n+1}(x) &=& -\frac{2}{\beta^S_0} P^{T,NS}_{S}(x)\otimes DA^{S,NS}_n(x).\nonumber \\
\eeqa
We construct  the boundary condition for the next stage of the evolution using the intermediate solution 
\beq
D_q(x,Q^2)=\sum_{n=0}^{n_0}\frac{DA^S_n(x)}{n!}
\left(\frac{\alpha({Q^2})}{\alpha({m_{2\lambda}}^2)}\right).
\eeq
evaluated at the next threshold $m_{2 \lambda}$. 
\beq
D_q(x,Q^2)=\sum_{n=0}^{n_0}\frac{DA^S_n(x)}{n!}
\log\left(\frac{\alpha({Q^2})}{\alpha({m_{2\lambda}}^2)}\right).
\eeq
The final solution is constructed using the recursion relations 
\beqa
DA^{ES}_0(x) &=& U(x)\otimes q(x,m_{ \tilde{q}}) \nonumber \\
DA^{ES}_{n+1}(x) &=& -\frac{2}{\beta^{ES}}P^{NS}_{ES}(x)\otimes DA^{ES}_n(x) \nonumber \\
\eeqa 

The final solution is written as 
\beqa
D_{q^{NS}}(x,Q^2) &=& D_{q^{NS}}(x,Q_0^2) + \sum_{n=1}^{n_0}
\frac{DA_n(x)}{n!}\log\left(\frac{\alpha({m_{2\lambda}}^2)}{\alpha(Q_0^2)}\right)
+ 
\sum_{n=1}^{n_0}\frac{DA^{S}_n(x)}{n!}\log\left(\frac{\alpha^{S}(m_{\tilde{q}}^2)}
{\alpha^{S}({m_{2\lambda}}^2)}\right) \nonumber \\
& + & \sum_{n=1}^{n_0}\frac{DA^{ES}_n(x)}{n!}\log\left(\frac{\alpha^{ES}(Q_f^2)}
{\alpha^{ES}({m_{\tilde{q}}}^2)}\right) \nonumber \\ 
\eeqa
As we have already mentioned above, in the analysis presented below the two susy scales 
will be collapsed into one ($m_{2\lambda}$).

\section{QCD Evolution of the Fragmentation Functions}
As we have already mentioned, the timelike and the spacelike evolution, in leading order, are essentially the same. The singlet kernels 
are just the transposed of the kernels describing the evolution of the 
parton distributions. The ordinary QCD kernels, in leading order, are given by

\beqa
P^{(0)}_{qq, NS} &=& C_F\left(\frac{1 + x^2}{1-x}\right)_+\nonumber \\
&=& C_F\left( \frac{2}{(1-x)_+}- 1 - x + \frac{3}{2}\,\delta(1-x)\right) \nonumber \\
P^{(0)}_{qq}(x)&=& P^{(0)}_{qq,NS}\nonumber \\
P^{(0)}_{qg}(x)&=&  \,\,n_f \left( x^2 + (1-x)^2\right)\nonumber \\
P^{(0)}_{gq}(x)&=& C_F {1 + (1-x)^2\over x}\nonumber \\
P^{(0)}_{gg}(x)&=& 2 N_c\left( {1\over (1-x)_+} + {1\over x}
-2 + x (1-x)\right) + {\beta_{o}\over 2}\delta(1-x)
\eeqa
with $\beta_0=11/3 \,C_A -4/3\, T_R\, n_f$ being 
the first coefficient of the QCD $\beta$-function and $T_R=1/2$. $n_f$ is the numbers of 
flavours.
The equations for the fragmentation functions in QCD are given by 
\beqa
\frac{d}{d \log(Q^2)}D^h_{q_i}(x,Q^2) 
&=&\frac{\alpha(Q^2)}{2 \pi}
\left( P_{qq}\otimes D^h_{q_i}+ \frac{1}{2 n_f}P_{g q}\otimes D^h_g\right)     \nonumber \\
\frac{d}{d \log(Q^2)}D^h_{g}(x,Q^2) 
&=&\frac{\alpha(Q^2)}{2\pi}
\left( P_{qg}\otimes \sum_i\left(D^h_{q_i}+D^h_{\bar{q}_i}\right) 
+ P_{g g}\otimes D^h_g \right)\nonumber \\
\eeqa
and, as usual, can be decomposed into a non-singlet and a singlet sector

\beqa  
Q^2 {d\over d Q^2} D^h_{q_i^{(-)}}(x, Q^2) &=& 
{\alpha(Q^2)\over 2 \pi} P_{qq,\,NS}(x, \alpha(Q^2))\otimes D^h_{q_{i}^{(-)}}(x, Q^2)
\eeqa
for the non-singlet distributions and 
\beqa
&&  { d\over d \log(Q^2)} \left( \begin{array}{c}
D^h_{q^{(+)}}(x, Q^2) \\
D^h_g(x, Q^2) \end{array}\right)=
\left(\begin{array}{cc}
P_{qq} & P_{gq} \\
P_{qg} & P_{gg}
\end{array} \right)\otimes 
\left( \begin{array}{c}
D^h_{q^{(+)}}(x,Q^2) \\
D^h_g(x,Q^2) \end{array}\right) \nonumber \\
\eeqa
for the singlet sector, where 

\beq
D^h_{q^{(-)}_i}(x,Q^2)=D^h_{q_i}- D^h_{\bar{q}_i}. 
\eeq
A fast strategy to solve these equations, as discussed in \cite{CC1}, 
where a complete leading order evolution has been implemented, is to solve the recursion relations 
numerically. In the timelike case -that we are considering- these relations are obtained 
as linear combinations of the spacelike ones. We get

\beqa
DA_{n+1}^{q_i^{(-)}}&=&-4 \frac{C_F}{\beta_0}\int_x^1 \frac{dy}{y}
\frac{y DA_n^{q_i^{(-)}}(y) - x DA_n^{q_i^{(-)}}(x) }{y-x} -
4 \frac{C_F}{\beta_0} \log(1-x)DA_n^{q_i^{(-)}}(x)\nonumber \\
&& + 2\frac{C_F}{\beta_0}\int_x^1\frac{dy}{y}\left(1 + z\right)DA_n^{q_i^{(-)}}(y) - 
3\frac{C_F}{\beta_0}DA_n^{q_i^{(-)}}(x)
\nonumber \\
DA_{n+1}^{q^{(-)}}&=&-4 \frac{C_F}{\beta_0}\int_x^1 \frac{dy}{y}
\frac{y DA_n^{q^{(-)}}(y) - x DA_n^{q^{(-)}}(x) }{y-x} -
4 \frac{C_F}{\beta_0} \log(1-x)DA_n^{q^{(-)}}(x)\nonumber \\
&& + 2\frac{C_F}{\beta_0}\int_x^1\frac{dy}{y}\left(1 + z\right)DA_n^{q^{(-)}}(y) - 
3\frac{C_F}{\beta_0}DA_n^{q^{(-)}}(x). \nonumber \\
\eeqa
A similar expansion is set up for the non-singlet variable 
$\chi_i=q_i^{(+)} - 1/n_f\,\, q^{(+)}$
\beqa
DA_{n+1}^{\chi_i}&=&-4 \frac{C_F}{\beta_0}\int_x^1 \frac{dy}{y}
\frac{y DA_n^{\chi_i}(y) - x DA_n^{\chi_i}(x) }{y-x} -
4 \frac{C_F}{\beta_0} \log(1-x)DA_n^{\chi_i}(x)\nonumber \\
&& + 2\frac{C_F}{\beta_0}\int_x^1\frac{dy}{y}\left(1 + z\right)DA_n^{\chi_i}(y) - 
3\frac{C_F}{\beta_0}DA_n^{\chi_i}(x). \nonumber \\
\eeqa
Similarly, the singlet equations generate recursion relations of the form

\beqa
DA_{n+1}^{q^{(+)}}(x) &=& -4\frac{C_F}{\beta_0}\int_x^1\frac{dy}{y}
\frac{y DA_n^{q^{(+)}}(y) - x DA_n^{q^{(+)}}(x)}{y-x} - 4 \frac{C_F}{\beta_0}\log(1-x) 
DA_n^{q^{(+)}}(x)\nonumber \\
&& + 2\frac{C_F}{\beta_0}\int_x^1\frac{dy}{y}\left( 1+ z\right)DA_n^{q^{(+)}}(y)
-3\frac{C_F}{\beta_0}DA_n^{q^{(+)}}(x)
\nonumber \\
&& -2\frac{C_F}{\beta_0 }\int_x^1\frac{dy}{y}
\frac{1 +(1-z)^2}{z}DA_n^g(y)
\nonumber \\
DA_{n+1}^g(x) &=&-4 \frac{C_A}{\beta_0}\int_x^1\frac{dy}{y}
\frac{y DA_n^{g}(y) - x DA_n^{g}(x)}{y-x} - 4\frac{C_A}{\beta_0}
\log(1-x)DA_n^g(x)\nonumber \\
&& - 2\frac{n_f}{\beta_0}\int_x^1\frac{dy}{y}\left( z^2 +(1-z)^2\right)DA_n^{q^{(+)}}(y)
- D A_n^g(x)\nonumber \\
&& -4 \frac{C_A}{\beta_0}\int_x^1\frac{dy}{y}
\left( \frac{1}{z} -2 +z(1-z)\right)DA_n^g(y).
\eeqa

The coefficients of the various flavours both for $D^h_{q_i}$ and $D^h_{\bar{q}_i}$ 
are then obtained from the relations 

\beqa
DA_n^{q_i}&=&\frac{1}{2}\left( DA_n^{\chi_i} + \frac{1}{n_f}DA_n^{q_i^{(-)}}\right)
\nonumber \\
DA_n^{\bar{q}_i}&=&\frac{1}{2}\left( DA_n^{\chi_i} - \frac{1}{n_f}DA_n^{q_i^{(-)}}\right)
\eeqa

\section{ The ESAP ($N=1$ QCD) Fragmentation}

Moving to $N=1$ QCD, we introduce recursion relations for appropriate 
linear combinations of non-singlet fragmentation functions 

\beqa
DA_{n+1}^{\chi_i} &=& -4\frac{C_F}{\beta_0}\int_x^1\frac{dy}{y}
\frac{y DA_n^{\chi_i}(y) - x DA_n^{\chi_i}(x)}{y-x} 
- 4 \frac{C_F}{\beta_0}\log(1-x) DA_n^{\chi_i}(x) 
\nonumber \\
&& - 2\frac{C_F}{\beta_0}DA_n^{\chi_i}(x)
- 2\frac{C_F}{\beta_0}\int_x^1\frac{dy}{y}z DA_n^{\tilde{\chi}_i}(y)
+2\frac{C_F}{\beta_o}
\int_x^1\frac{dy}{y}\left(1 +z\right)D A_n^{\chi_i}(y) \nonumber \\
\eeqa

\beqa
DA_{n+1}^{\tilde{\chi}_i} &=& -4\frac{C_F}{\beta_0}\int_x^1\frac{dy}{y}
\frac{y DA_n^{\tilde{\chi}_i}(y) - x DA_n^{\tilde{\chi}_i}(x)}{y-x} 
- 4 \frac{C_F}{\beta_0}\log(1-x) DA_n^{\tilde{\chi}_i}(x) \nonumber \\
&& - 2\frac{C_F}{\beta_0}DA_n^{\tilde{\chi}_i}(x)
- 2\frac{C_F}{\beta_0}\int_x^1\frac{dy}{y}z DA_n^{\chi_i}(y) 
+2\frac{C_F}{\beta_o}
\int_x^1\frac{dy}{y}\left(1 +z\right)D A_n^{\tilde{\chi_i}}(y)\nonumber \\
\eeqa

and two similar non-singlet equations for $D^h_{q_i^{(-)}}$ and $D^h_{\tilde{q}_i^{(-)}}$

\beqa
DA_{n+1}^{q_i^{(-)}} &=& -4\frac{C_F}{\beta_0}\int_x^1\frac{dy}{y}
\frac{y DA_n^{q^{(-)}_i}(y) - x DA_n^{q^{(-)}_i}(x)}{y-x} 
- 4 \frac{C_F}{\beta_0}\log(1-x) DA_n^{q^{(-)}_i}(x) \nonumber \\
&& - 2\frac{C_F}{\beta_0}DA_n^{q^{(-)}_i}(x)
- 2\frac{C_F}{\beta_0}\int_x^1\frac{dy}{y}z DA_n^{\tilde{q}^{(-)}_i}(y)
+2\frac{C_F}{\beta_o}
\int_x^1\frac{dy}{y}\left(1 +z\right)D A_n^{q_i^{(-)}}(y)
 \nonumber \\
\label{one}
\eeqa

\beqa
DA_{n+1}^{\tilde{q}^{(-)}_i} &=& -4\frac{C_F}{\beta_0}\int_x^1\frac{dy}{y}
\frac{y DA_n^{\tilde{q}^{(-)}_i}(y) - x DA_n^{\tilde{q}^{(-)}_i}(x)}{y-x} 
- 4 \frac{C_F}{\beta_0}\log(1-x) DA_n^{\tilde{q}^{(-)}_i}(x) \nonumber \\
&& - 2\frac{C_F}{\beta_0}DA_n^{\tilde{q}^{(-)}_i}(x)
- 2\frac{C_F}{\beta_0}\int_x^1\frac{dy}{y}z DA_n^{q^{(-)}_i}(y) 
+2\frac{C_F}{\beta_o}
\int_x^1\frac{dy}{y}\left(1 +z\right)D A_n^{\tilde{q}_i^{(-)}}(y).\nonumber \\
\label{two}
\eeqa

Analogous equations are satisfied by the combinations 
$D^h_{q^{(-)}}$ and $D^h_{\tilde{q}^{(-)}}$ by replacing in (\ref{one}) and (\ref{two})
$D^h_{q_i^{(-)}}\to D^h_{q^{(-)}}$ and $D^h_{\tilde{q}_i^{(-)}} \to D^h_{\tilde{q}^{(-)}}$.

Moving to the singlet sector we obtain
\beqa
DA_{n+1}^g(x) &=& - 4 \frac{C_A}{\beta_0}\int_x^1 \frac{dy}{y}	
\frac{y DA_n^g(y) - x DA_n^g (x)}{ y -x}
- 4 \frac{C_A}{\beta_0}\log(1-x)DA_n^g(x) -D A_n^g(x)
\nonumber \\
&& + 2 \frac{C_A}{\beta_0}\int_x^1\frac{dy}{y}\left( 1 + z\right)DA_n^g(y)
 -2 \frac{C_A}{\beta_0}\int_x^1\frac{dy}{y}\left( 
\frac{2}{z} + z - 2\right)DA_n^g(y)\nonumber \\
&& +  2 \frac{C_A}{\beta_0}\int_x^1\frac{dy}{y}\left(z^2 + (1-z)^2\right)
DA_n^g(y)
 -2 \frac{n_f}{\beta_0}\int_x^1\frac{dy}{y}
\left( z^2 +(1-z)^2\right) DA_n^q(y)\nonumber \\
&& -2 \frac{C_A}{\beta_0}\int_x^1\frac{dy}{y}\left(z^2 + (1-z)^2\right)DA_n^\lambda(y)
-4\frac{n_f}{\beta_0}\int_x^1\frac{dy}{y}\, z\left( 1-z\right)
DA_n^{\tilde{q}}(y) \nonumber \\
\eeqa

\beqa
DA_n^{\lambda}(x) &=& -2\frac{C_A}{\beta_0}\int_x^1
\frac{dy}{y}\left(\frac{2}{z} + z - 2\right)DA_n^g(y)
-4 \frac{C_A}{\beta_0}\int_x^1\frac{dy}{y}
\frac{DA_n^\lambda(y)- x DA_n^\lambda(x)}{y-x}\nonumber \\
&& -4 \frac{C_A}{\beta_0}\log(1-x)DA_n^\lambda(x)
+2 \frac{C_A}{\beta_0}\int_x^1 \frac{dy}{y}
\left( 1 +z\right)DA_n^\lambda(y) - DA_n^\lambda(x)
\nonumber \\
&& -2\frac{n_f}{\beta_0}\int_x^1\frac{dy}{y}\left( 1-z \right)DA_n^q(y)
-2\frac{n_f}{\beta_0}\int_x^1\frac{dy}{y} z DA_n^{\tilde{q}}(y)
\eeqa

\beqa
DA_{n+1}^{q^{(+)}}(x) &=& -2\frac{C_F}{\beta_0}\int_x^1 
\frac{dy}{y}\left(\frac{2}{z} + z -2\right)DA_n^g(y) 
-2\frac{C_F}{\beta_0}\int_x^1\frac{dy}{y}\left( 1-z \right) 
DA_n^\lambda(y)\nonumber \\
&& -4 \frac{C_F}{\beta_0}\int_x^1\frac{dy}{y}
\frac{y DA_n^{q^{(+)}}(y) - xDA_n^{q^{(+)}}(x)}{y-x} - 
4\frac{C_F}{\beta_0} \log(1-x) DA_n^{q^{(+)}}(x)
\nonumber \\
&& 
+ 2 \frac{C_F}{\beta_0}\int_x^1 \frac{dy}{y}
\left( 1 + z\right)DA_n^{q^{(+)}}(y) -2\frac{C_F}{\beta_0}DA_n^{q^{(+)}}(x)
-2\frac{C_F}{\beta_0}\int_x^1\frac{dy}{y}z DA_n^{\tilde{q}^{(+)}}(y)
\nonumber \\
\eeqa

\beqa
DA_{n+1}^{\tilde{q}^{(+)}}(x)&=&-2\frac{C_F}{\beta_0}\int_x^1
\frac{dy}{y}\left(\frac{2}{z}-2\right)DA_n^g(y)
-2\frac{C_F}{\beta_0}\int_x^1\frac{dy}{y}\left(DA_n^\lambda(x) -DA_n^{\tilde{q}^{(+)}}(y)\right)
\nonumber \\
&& -4\frac{C_F}{\beta_0}\int_x^1\frac{dy}{y}
\frac{y DA_n^{\tilde{q}}(y) - x DA_n^{\tilde{q}}(x)}{y-x}
+ 4\frac{C_F}{\beta_0}\int_x^1\frac{dy}{y}DA_n^{\tilde{q}}(y)\nonumber \\ 
&& -
4\frac{C_F}{\beta_0}\log(1-x)DA_n^{\tilde{q}^{(+)}}(y)
-2\frac{C_F}{\beta_0}DA_n^{\tilde{q}^{(+)}}(x)
\eeqa
where $z\equiv x/y$.

\section{Numerical Results}
As an illustration of the procedure we adopt in our studies, let's consider the decay 
of a hypothetical massive state of mass 1 TeV into supersymmetric partons. 
The decay can proceed, for instance, through a regular $q\bar{q}$ 
channel and a shower is developed starting from the quark pair. The $N=1$ DGLAP 
equation describes in the leading logarithmic approximation the evolution 
of the shower which accompanies the pair, and we are interested in studying the 
impact of the supersymmetry breaking scale ($m_\lambda$) on the fragmentation. 
In our runs we have chosen the initial set of Ref.~\cite{kkp}. 		
The parameterizations are shown in the appendix just 
for illustrative purposes. 
 In our analysis we focus on proton fragmentation 
functions. 
As in Ref.~\cite{kkp}, we introduce the scaling variable
\begin{equation}
\bar s=\ln\frac{\ln\left(\mu^2/\Lambda_{\overline{\mathrm{MS}}}^{(5)}\right)}
{\ln\left(\mu_0^2/\Lambda_{\overline{\mathrm{MS}}}^{(5)}\right)}.
\end{equation}
In LO  we have $\Lambda_{\overline{\mathrm{MS}}}^{(5)}=88$~MeV 
\cite{kkp}.
We use three different values for $\mu_0$, namely \cite{kkp}
\begin{eqnarray}
\mu_0&=&\left\{\begin{array}{l@{\quad\mbox{if}\quad}l}
\sqrt2~\mbox{GeV}, & a=u,d,s,g\\
m(\eta_c)=2.9788~\mbox{GeV}, & a=c\\
m(\Upsilon)=9.46037~\mbox{GeV}, & a=b
\end{array}\right..
\end{eqnarray}
This leads to three different definitions of $\bar s$.
For definiteness, we use the symbol $\bar s_c$ for charm and $\bar s_b$ for
bottom along with $\bar s$ for the residual partons.
We parameterize the f.f.'s as
\begin{equation}
\label{temp}
D(x,\mu^2)=Nx^\alpha(1-x)^\beta\left(1+\frac{\gamma}{x}\right)
\end{equation}
and express the coefficients $N$, $\alpha$, $\beta$, and $\gamma$ as
polynomials in $\bar s$, $\bar s_c$, and $\bar s_b$.
For $\bar s=\bar s_c=\bar s_b=0$, the parameterizations agree with Eq.~(2) of
Ref.~\cite{kkp} in combination with the appropriate entries in Table~2 of that
paper.
The charm and bottom parameterizations must be put to zero by hand for
$\bar s_c<0$ and $\bar s_b<0$, respectively.

Typical fragmentation functions in QCD involve final states with 
$p$, $\bar{p}$, $\pi^{\pm},\pi^{0}$ and kaons $k^{\pm}$.
We have included the relevant sets from ref.~\cite{kkp} 
that we have employed 
in our numerical analysis in an appendix. 
We refer to the original literature for a list of all the fragmentation sets. 
 We have chosen 
an initial evolution scale of $10$ GeV and varied both the mass of the susy partners 
(we assume for simplicity that these are all degenerate) 
and the final evolution scale. Since our concern is in establishing the impact of 
supersymmetric evolution and compare it to standard QCD evolution across large evolution intervals, we plot the initial fragmentation functions, 
the regularly evolved QCD functions and the SQCD/QCD evolved ones. 
The latter two are originated from the same low energy form of Ref.~\cite{kkp}.
The figures show that the effects of supersymmetric evolution 
are small within the range described by the factorization scales $Q_f$ and 
$Q_i$ ($Q_f=10^3$ GeV, $Q_i=200$ GeV). 
We mention that $Q_f$ is the starting scale (the highest scale) 
at which the decay of the supersymmetric partons starts. $Q_i$ 
is fixed by the gluino/squark masses and coincides with them. 
It is possible to include in the evolution of the fragmentation 
functions also different thresholds associated 
with more complex spectra in which the susy partners are not degenerate. 
These effects are negligible. Threshold enhancements 
may require a more accurate treatment and will be discussed elsewhere, 
however we don't expect them to play any important role, 
especially since we are interested in very extended renormalization group runnings.  
Fig.~2 shows the initial condition for the up quark fragmentation function. 
We have chosen $Q_0=10$ GeV, extrapolated from collider data in Ref.~\cite{kkp} 
in their global analysis. The evolution of this function follows QCD 
from this lowest scale up to a scale of $400$ GeV $(m_{2\lambda})$. 
Above this scale we use the full $N=1$ evolution. The initial fragmentation scale, here denoted by $Q_f$ is 1 TeV. In general, the small-x (diffractive) 
region gets slightly enhanced. The highest scale is not large enough to allow a 
discrimination of susy effects from the non supersymmetric ones. 
The trend of the two evolved curves is similar. In Fig.~3 an analogous behaviour 
is found for the fragmentation functions of charm and strange quarks. Again, 
small-x enhancements are seen, but regular and susy evolution are hardly 
distinguishable. The situation appears to be completely different 
for the gluon fragmentation functions (f.f's) (Fig.~4). The regular and the SQCD evolved 
f.f.'s differ largely in the diffractive region, and this clearly will show up in the 
spectrum of the primary protons if the decaying state has a supersymmetric content. 
As we raise the final evolution scale 
we start seeing more pronounced differences between regular 
and supersymmetric distributions. 
We have shown in Fig.~5 the squark f.f.'s for all the flavours and 
the one of the gluino for comparison. The scalar charm distribution appear 
to grow slightly faster then the remaining scalar ones. The gluino f.f. 
is still the fastest growing at small-x values.

 In Figs.~6 and 7 this trend is quite evident for the quark  and gluon fragmentation functions respectively. The latter is the distributions which remains by far the most sensitive to changes in the highest fragmentation scale. It is therefore very likely that the spectrum of final protons is substantially modified 
by this fast modifications of the gluon density. A sustained stability -as we raise the 
supersymmetric scale - of all the distributions is also noticed. 
In fact, in Fig.~ 8 we vary the gluino mass 
for a fixed final highest scale and notice a mild variation of the fragmentation 
function of the up quark. A similar behaviour is recorded in Fig.~9, 
where we compare the fragmentation functions of the up quark and of its superpartner 
at various $m_\lambda$'s. The shape of the gluino density 
as a function of the the initial fragmentation scale is shown in Fig.~10. 
The growth appears to be fast and is rather pronounced. 
Fig.~11 finally illustrates the behaviour of f.f's of all types at a large initial 
fragmentation scale. In order to illustrate the different behaviours 
that the quark/squark sector has compared to the gluon/gluino sector, 
we have shown the f.f.'s for the $b$ quark and the $\tilde{b}$ squark. 
Although all the functions get a small-x enhancement or are more supported in this 
x-region, the largest fragmentation appears to come from the gluon-gluino sector. 

\section{Conclusions}
We have presented a first detailed study of the fragmentation functions 
of susy QCD in the leading logarithmic approximation and presented, 
for the first time, results for the f.f.'s 
of all the partons into protons within a radiatively generated model. 
Our analysis is motivated by the growing interest that high energy 
cosmic rays experiments will be receiving in the next several years 
from the astroparticle/high energy physics community. There are hints, 
from this work, that supersymmetric effects may affect substantially 
the fragmentation region and the spectra of the hadronic component 
of cosmic rays which is expected to play a dominant role as one approaches the GZK cutoff. 
Our analysis indicates that fragmentation probabilities are redistributed 
among the various open channel and have an impact 
on the spectrum of the primaries. Compared to initial 
state scaling violations where only a gluino of small mass seems to 
enhance the radiative structure of parton distributions \cite{CC1,CC2,CC3}, the 
cosmic ray story clearly points to another direction. Given the very 
large energy available in the fragmentation region, and assuming 
a supersymmetry breaking scale of the order of 1 TeV, the f.f's, 
which control the multiplicities of the hadron component of the various channels are substantially affected by the presence of a supersymmetric 
scale at the highest energy. The natural question to ask is then: 
what are the distributions of supersymmetric partons in a decaying metastable state 
of very large mass prior to fragmentation? At such large scales, 
so far from the 1 TeV scale preferred by many standard supersymmetric arguments,
the Renormalization Group strategy 
can be easily embraced to its fullest extent with sizeable consequences 
on the low energy end. On this and other related issues 
we hope to return in more detail in the near future.     

\vspace{1cm}
\centerline{\bf Acknowledgements}
We thank S. Sarkar and R. Toldra for discussions. 
C.C. and A.F. thank respectively
the Theory Groups at Oxford and Lecce for hospitality. 
The work of C.C. is supported in part by INFN 
(iniziativa specifica BARI-21) and by MURST. The work of A.F. is supported
by PPARC.    

\section{Appendix 1. Input Functions}
For convenience we have included below the list 
of f.f.'s taken from Ref.~\cite{kkp} that we have used

\begin{itemize}
\item $D_u^{p/\bar p}(x,\mu^2)=2D_d^{p/\bar p}(x,\mu^2)$:
\begin{eqnarray}
N&=&0.40211-0.21633\bar s-0.07045\bar s^2+0.07831\bar s^3\nonumber\\
\alpha&=&-0.85973+0.13987\bar s-0.82412\bar s^2+0.43114\bar s^3\nonumber\\
\beta&=&2.80160+0.78923\bar s-0.05344\bar s^2+0.01460\bar s^3\nonumber\\
\gamma&=& 0.05198\bar s-0.04623\bar s^2
\end{eqnarray}
\item $D_s^{p/\bar p}(x,\mu^2)$:
\begin{eqnarray}
N&=&4.07885-2.97392\bar s-0.92973\bar s^2+1.23517\bar s^3\nonumber\\
\alpha&=&-0.09735+0.25834\bar s-1.52246\bar s^2+0.77060\bar s^3\nonumber\\
\beta&=&4.99191+1.14379\bar s-0.85320\bar s^2+0.45607\bar s^3\nonumber\\
\gamma&=&0.07174 \bar s-0.08321\bar s^2
\end{eqnarray}
\item $D_c^{p/\bar p}(x,\mu^2)$:
\begin{eqnarray}
N&=&0.11061-0.07726\bar s_c+0.05422\bar s_c^2-0.03364\bar s_c^3\nonumber\\
\alpha&=&-1.54340-0.20804\bar s_c+0.29038\bar s_c^2-0.23662\bar s_c^3
\nonumber\\
\beta&=&2.20681+0.62274\bar s_c+0.29713\bar s_c^2-0.21861\bar s_c^3\nonumber\\
\gamma&=&0.00831\bar s_c+0.00065\bar s_c^2
\end{eqnarray}
\item $D_b^{p/\bar p}(x,\mu^2)$:
\begin{eqnarray}
N&=&40.0971-123.531\bar s_b+128.666\bar s_b^2-29.1808\bar s_b^3\nonumber\\
\alpha&=&0.74249-1.29639\bar s_b-3.65003\bar s_b^2+3.05340\bar s_b^3
\nonumber\\
\beta&=&12.3729-1.04932\bar s_b+0.34662\bar s_b^2-1.34412\bar s_b^3\nonumber\\
\gamma&=&-0.04290\bar s_b-0.30359\bar s_b^2
\end{eqnarray}
\item $D_g^{p/\bar p}(x,\mu^2)$:
\begin{eqnarray}
N&=&0.73953-1.64519\bar s+1.01189\bar s^2-0.10175\bar s^3\nonumber\\
\alpha&=&-0.76986-3.58787\bar s+13.8025\bar s^2-13.8902\bar s^3\nonumber\\
\beta&=&7.69079-2.84470\bar s-0.36719\bar s^2-2.21825\bar s^3\nonumber\\
\gamma&=&1.26515\bar s-1.96117\bar s^2
\end{eqnarray}
\end{itemize}

\section{Appendix 2. The Weights of the N=1 Kernels}
We briefly recall the numerical strategy 
employed in this analysis. A more detailed description will be given elsewhere. 
We just mention that the 
radiative generation of supersymmetric distributions requires 
special accuracy since these scaling violations grow up very slowly. 
We define $\bar{P}(x)\equiv x P(x)$ and $\bar{A}(x)\equiv x A(x)$. 
We also define the convolution product  

\beq
J(x)\equiv\int_x^1 \frac{dy}{y}\left(\frac{x}{y}\right) P\left(\frac{x}{y}\right)\bar{A}(y). \ 
\eeq
The integration interval in $y$ at any fixed x-value is partitioned in an array of 
increasing points ordered from left to right 
$\left(x_0,x_1,x_2,...,x_n,x_{n+1}\right)$ 
with $x_0\equiv x$ and $x_{n+1}\equiv 1$ being the upper edge of the integration 
region. One constructs a rescaled array 
$\left(x,x/x_n,...,x/x_2,x/x_1, 1 \right)$. We define 
$s_i\equiv x/x_i$, and $s_{n+1}=x < s_n < s_{n-1}<... s_1 < s_0=1$.
We get 
\beq
J(x)=\sum_{i=0}^N\int_{x_i}^{x_{i+1}}\frac{dy}{y}
\left(\frac{x}{y}\right) P\left(\frac{x}{y}\right)\bar{A}(y) 
\eeq
At this point we introduce the linear interpolation 
\beq
\bar{A}(y)=\left( 1- \frac{y - x_i}{x_{i+1}- x_i}\right)\bar{A}(x_i) + 
\frac{y - x_i}{x_{i+1}-x_i}\bar{A}(x_{i+1})
\label{inter}
\eeq
and perform the integration on each subinterval with a change of variable $y->x/y$ and replace the integral $J(x)$ with 
its discrete approximation $J_N(x)$
to get 
\beqa
J_N(x) &=& \bar{A}(x_0)\frac{1}{1- s_1}\int_{s_1}^1 \frac{dy}{y}P(y)(y - s_1) \nonumber \\
&+& \sum_{i=1}^{N}\bar{A}(x_i) \frac{s_i}{s_i - s_{i+1}}
\int_{s_{i+1}}^{s_i} \frac{dy}{y}P(y)(y - s_{i+1})\nonumber \\
& -& \sum_{i=1}^{N}\bar{A}(x_i) \frac{s_i}{s_{i-1} - s_{i}}
\int_{s_{i}}^{s_{i-1}} \frac{dy}{y}P(y)(y - s_{i-1}) \nonumber \\
\eeqa
with the condition $\bar{A}(x_{N+1})=0$.
Introducing the coefficients  $W(x,x)$ and $W(x_i,x)$, the integral 
is cast in the form 
\beq
J_N(x)=W(x,x) \bar{A}(x) + \sum_{i=1}^{n} W(x_i,x)\bar{A}(x_i)  
\eeq
where
\beqa
W(x,x) &=& \frac{1}{1-s_1} \int_{s_1}^1 \frac{dy}{y}(y- s_1)P(y), \nonumber \\
W(x_i,x) &=& \frac{s_i}{s_i- s_{i+1}}
\int_{s_{i+1}}^{s_i} \frac{dy}{y}\left( y - s_{i+1}\right) P(y) \nonumber \\
& -& \frac{s_i}{s_{i-1} - s_i}\int_{s_i}^{s_{i-1}}\frac{dy}{y}\left(
y - s_{i-1}\right) P(y).\nonumber \\
\eeqa

We recall that 
\beq
  \int_0^1 dx \frac{f(x)}{(1-x)_+}=\int_0^1 {dy}\frac{f(y)- f(1)}{1-y}
\eeq
and that 
\beq
 \frac{1}{(1-x)_+}\otimes f(x)\equiv 
\int_x^1\frac{dy}{y}\frac{\,\,y f(y) - x f(x)}{y-x} + f(x)\log(1-x) 
\eeq
as can can be shown quite straightforwardly. 

We also introduce the expressions 

\beqa
In_0(x) & = & 
 \frac{1}{1- s_1} \log(s_1) + \log(1- s_1) \nonumber \\
\nonumber \\
Jn_i(x) & = & \frac{1}{s_i - s_{i +1}}
\left[ \log\left(\frac{1 - s_{i+1}}{1 - s_i}\right) 
+ s_{i+1} \log\left(\frac{1- s_i}{1 - s_{i+1}}\frac{s_{i+1}}{s_i}\right)\right]
\nonumber \\
Jnt_i(x) & = & \frac{1}{s_{i-1}- s_i}\left[ \log\left(\frac{1 - s_i}{1 - s_{i-1}}\right) 
+ s_{i-1}\log\left( \frac{s_i}{s_{i-1}}\right) + s_{i-1}\left( 
\frac{1 - s_{i-1}}{1 - s_i}\right)\right],   \,\,\,\,\,\ i=2,3,..N \nonumber \\
Jnt_1(x) &=& \frac{1}{1- s_1}\log s_1. \nonumber \\
\eeqa

Using the linear interpolation formula (\ref{inter})  we get the 
relation 

\beqa
\int_x^1\frac{dy}{y} \frac{ y A_n(y) - x A_n(x)}{y-x} &=&
- \log(1-x) A_n(x) + A_n(x) In_0(x)\nonumber \\
&&  + \sum_{i=1}^N A_n(x_i) \left( Jn_i(x) - Jnt_i(x)\right)
\eeqa

which has been used for a fast 
and accurate numerical implementation of the recursion relations.

\newpage

\begin{figure}
\centerline{\includegraphics[angle=0,width=.9\textwidth]{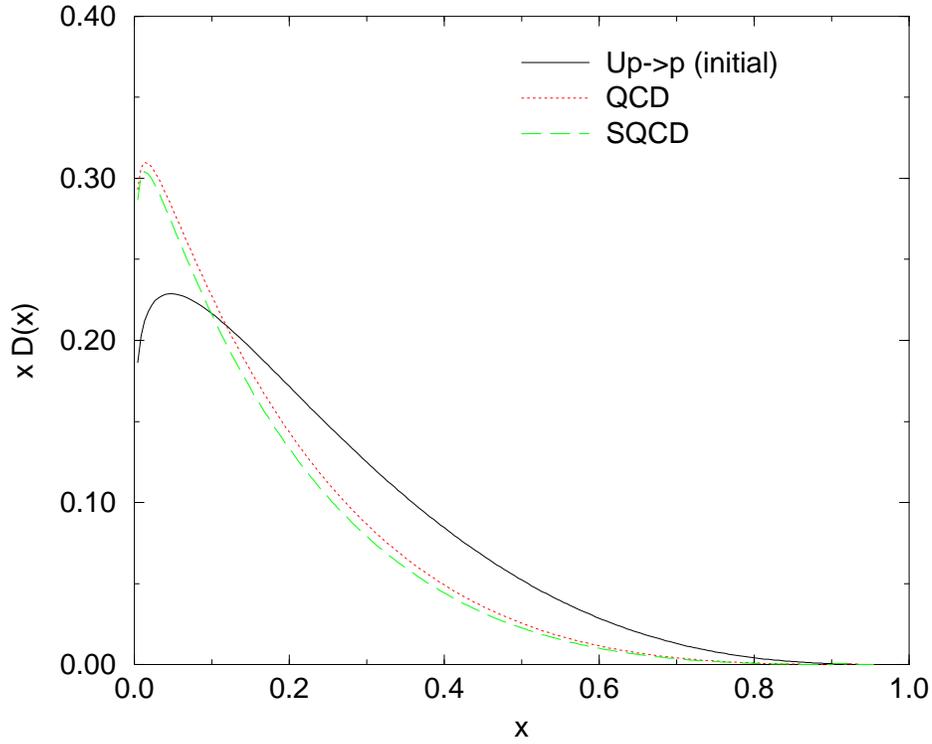}}
\centerline{\includegraphics[angle=0,width=.9\textwidth]{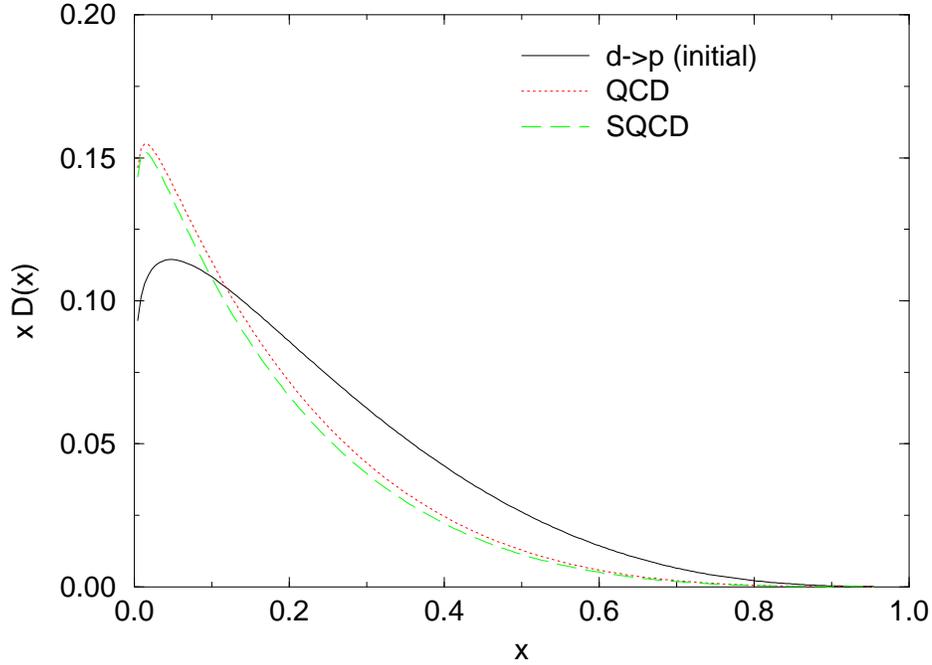}}
\caption{Fragmentation function for the 
up (top graph) and down  quark (bottom graph)  $x D_{u,d}^{p,\bar{p}}(x,Q^2)$  at the lowest scale (input)
$Q_0=10$ GeV, and their QCD (or regular) and SQCD/QCD evolutions 
with $Q_f=$$10^3$ GeV.The susy fragmentation scale is chosen to be $200$ GeV.  }.
\end{figure}

\begin{figure}
\centerline{\includegraphics[angle=0,width=.9\textwidth]{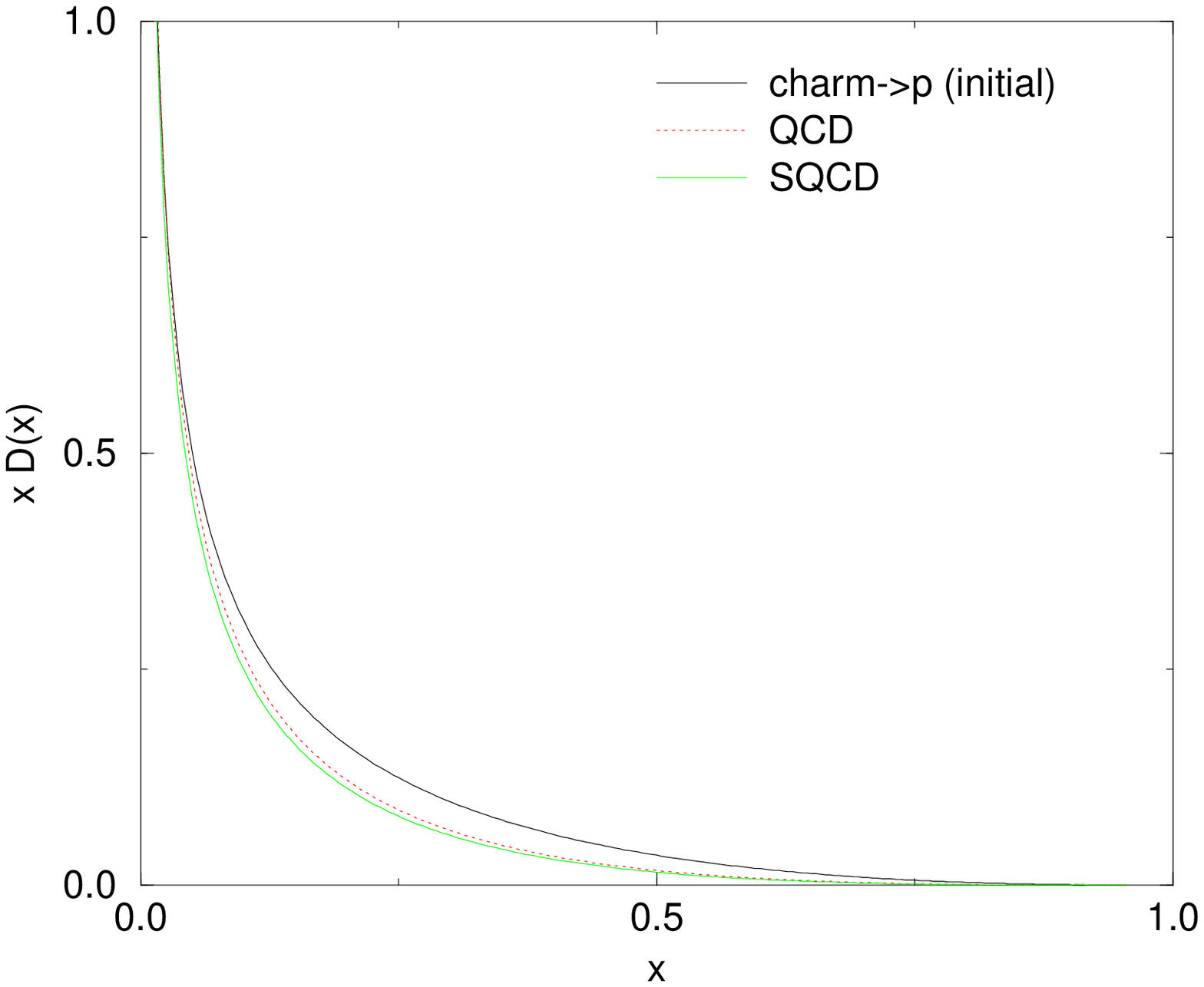}}
\centerline{\includegraphics[angle=0,width=.9\textwidth]{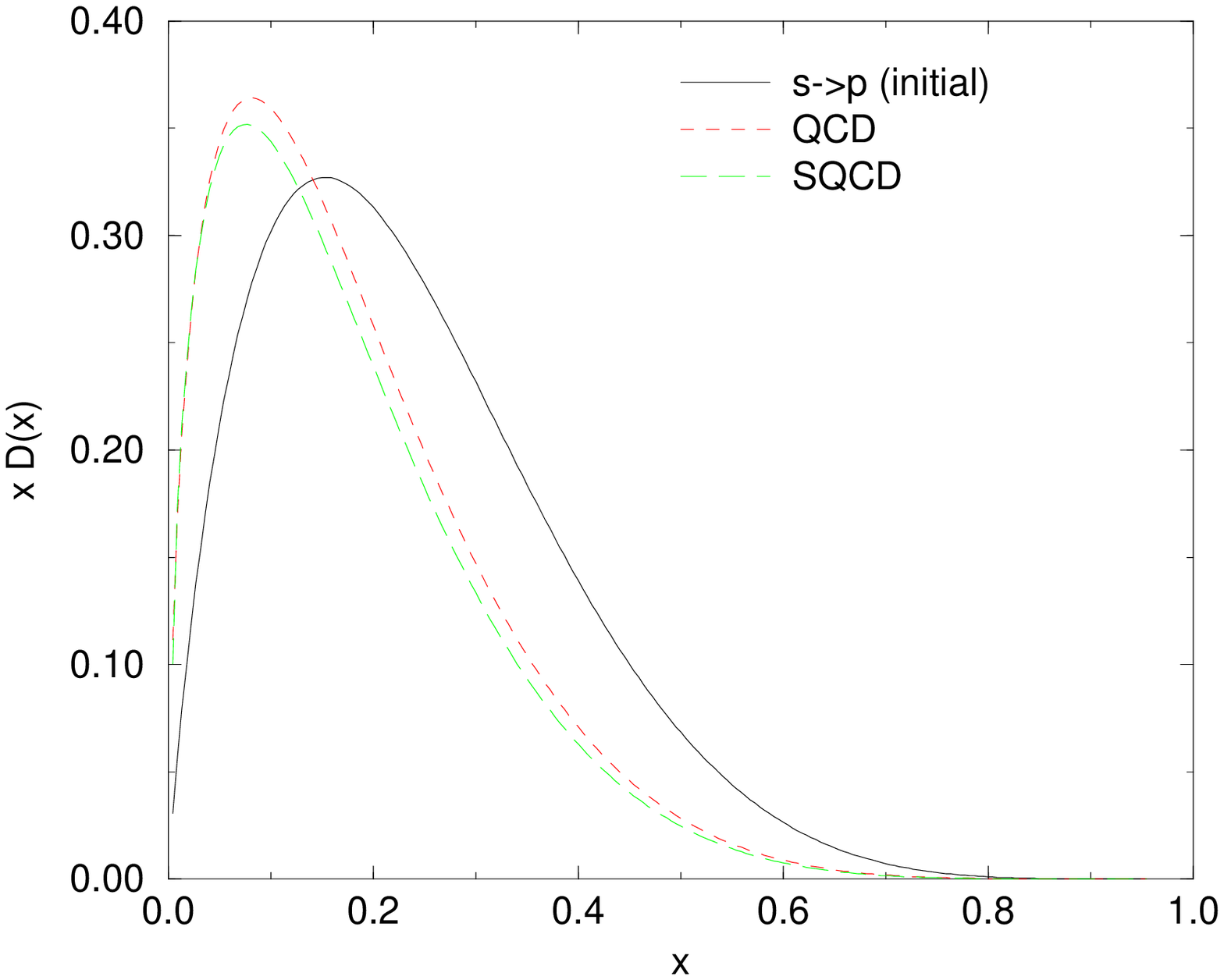}}
\caption{ Fragmentation functions into protons of 
the charm and strange quarks $x D_{s,c}^{p,\bar{p}}(x,Q^2)$  at the lowest scale (input)
$Q_0=10$ GeV, and its evolved QCD (regular) and SQCD/QCD evolutions 
with $Q_f=$$10^3$ GeV.The susy fragmentation scale is chosen to be $200$ GeV.}
\end{figure}

\begin{figure}
\centerline{\includegraphics[angle=0,width=1.1\textwidth]{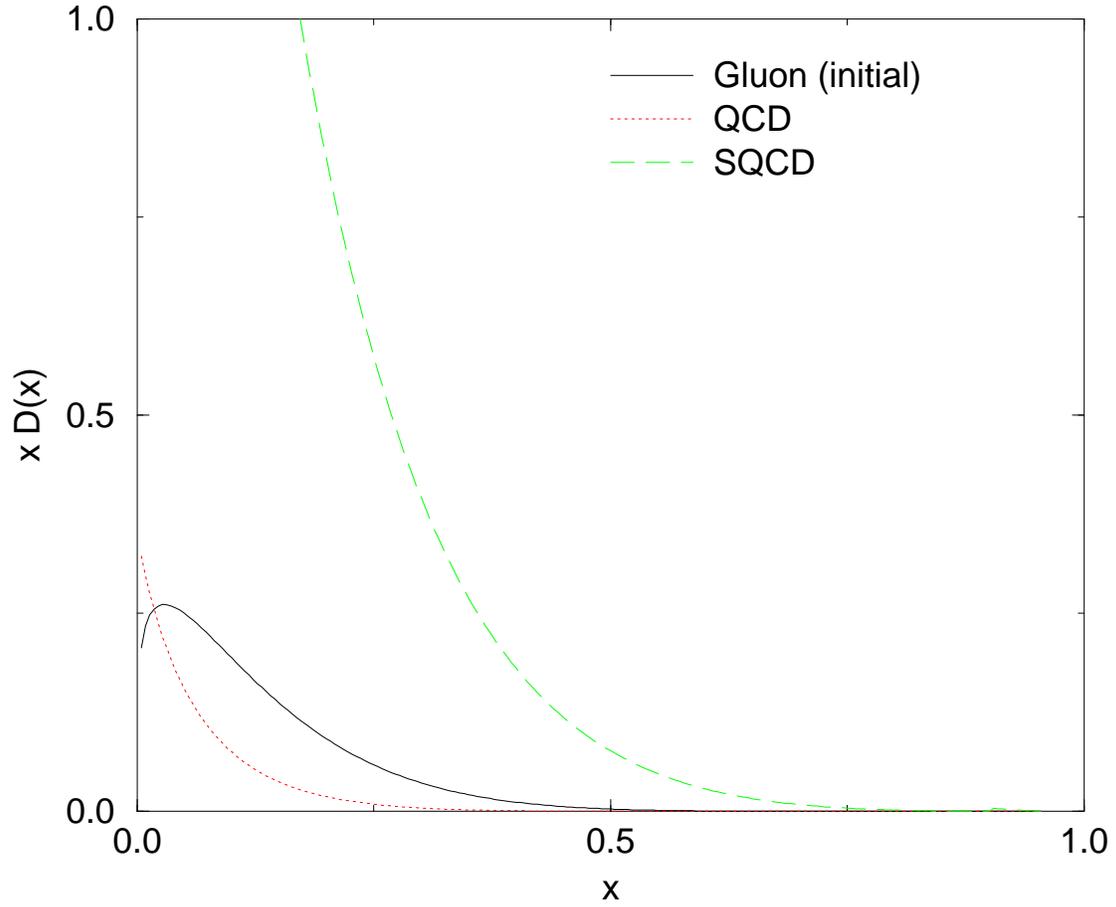}}
\caption{The gluon fragmentation function $x D_{g}^{p,\bar{p}}(x,Q^2)$  at the lowest scale (input)
$Q_0=10$ GeV, and its evolved QCD (regular) and SQCD/QCD evolutions 
with $Q_f=$$10^3$ GeV.The susy fragmentation scale is chosen to be $200$ GeV.}
\end{figure}
\begin{figure}
\centerline{\includegraphics[angle=0,width=1.1\textwidth]{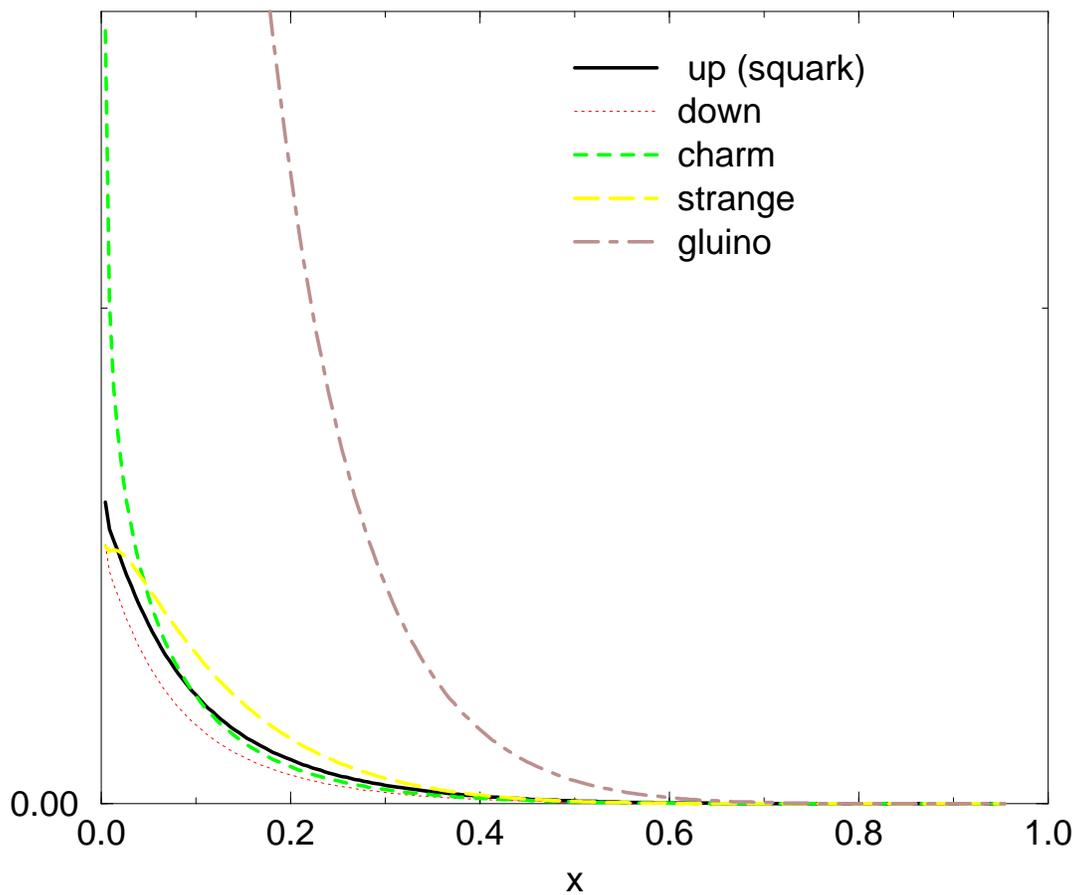}}
\caption{The fragmentation functions of squarks and gluino
 at the lowest scale (input)
$Q_0=10$ GeV, with $Q_f=10^3$ GeV and susy scale $200$ GeV.}
\end{figure}

\begin{figure}
\centerline{\includegraphics[angle=0,width=1.1\textwidth]{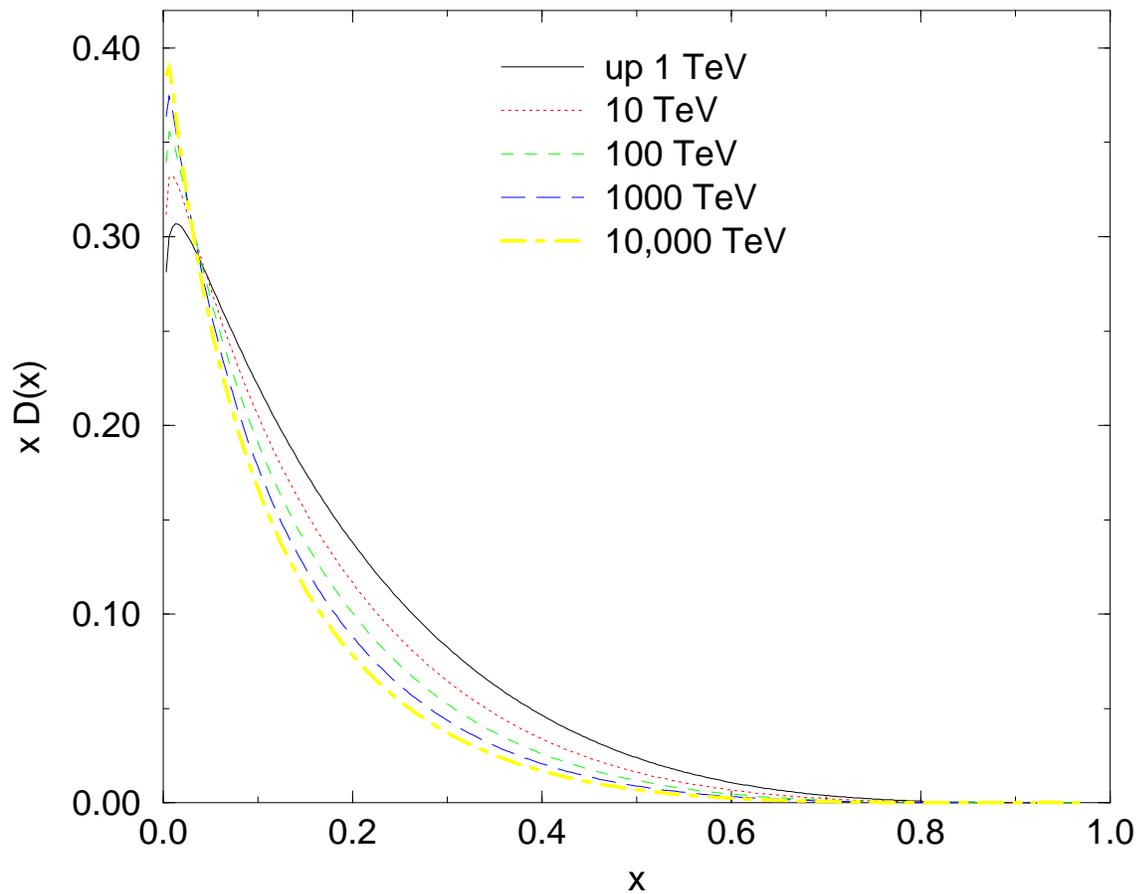}}
\caption{The up quark fragmentation function $x D_{u}^{p,\bar{p}}(x,Q^2)$ 
evolved with SQCD/QCD for varying final values of $Q_f$. 
The susy fragmentation scale is chosen to be $200$ GeV.}
\end{figure}

\begin{figure}
\centerline{\includegraphics[angle=0,width=1.1\textwidth]{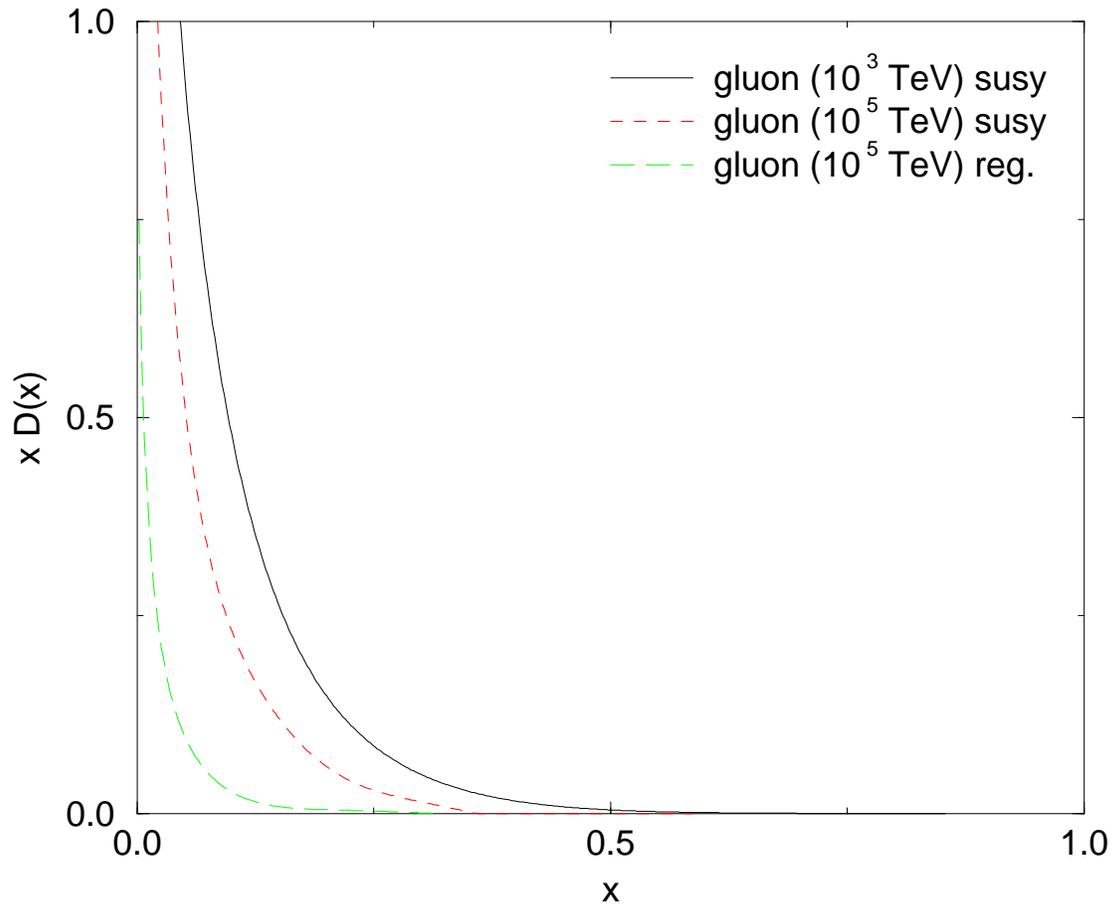}}
\caption{The gluon fragmentation function $x D_{g}^{p,\bar{p}}(x,Q^2)$ 
evolved regularly and according to SQCD/QCD for varying final values of $Q_f$. 
The susy fragmentation scale is chosen to be $200$ GeV.}
\end{figure}

\begin{figure}
\centerline{\includegraphics[angle=0,width=1.1\textwidth]{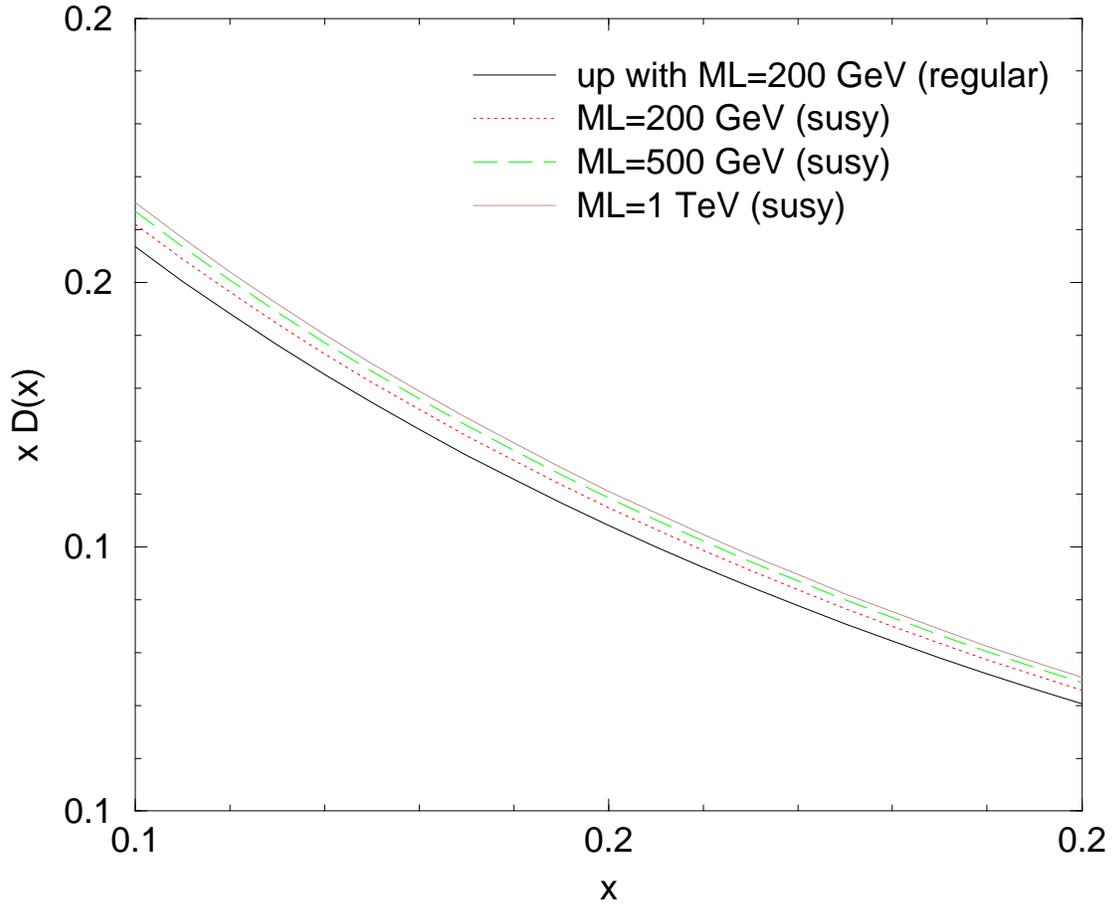}}
\caption{The up quark fragmentation function $x D_{u}^{p,\bar{p}}(x,Q^2)$ 
evolved regularly and according to SQCD/QCD for varying final values of the susy scale $m_{2\lambda}$  and $Q_f=10^8$ GeV. 
The susy fragmentation scale is chosen to be $200$ GeV.}
\end{figure}

\begin{figure}
\centerline{\includegraphics[angle=0,width=1.1\textwidth]{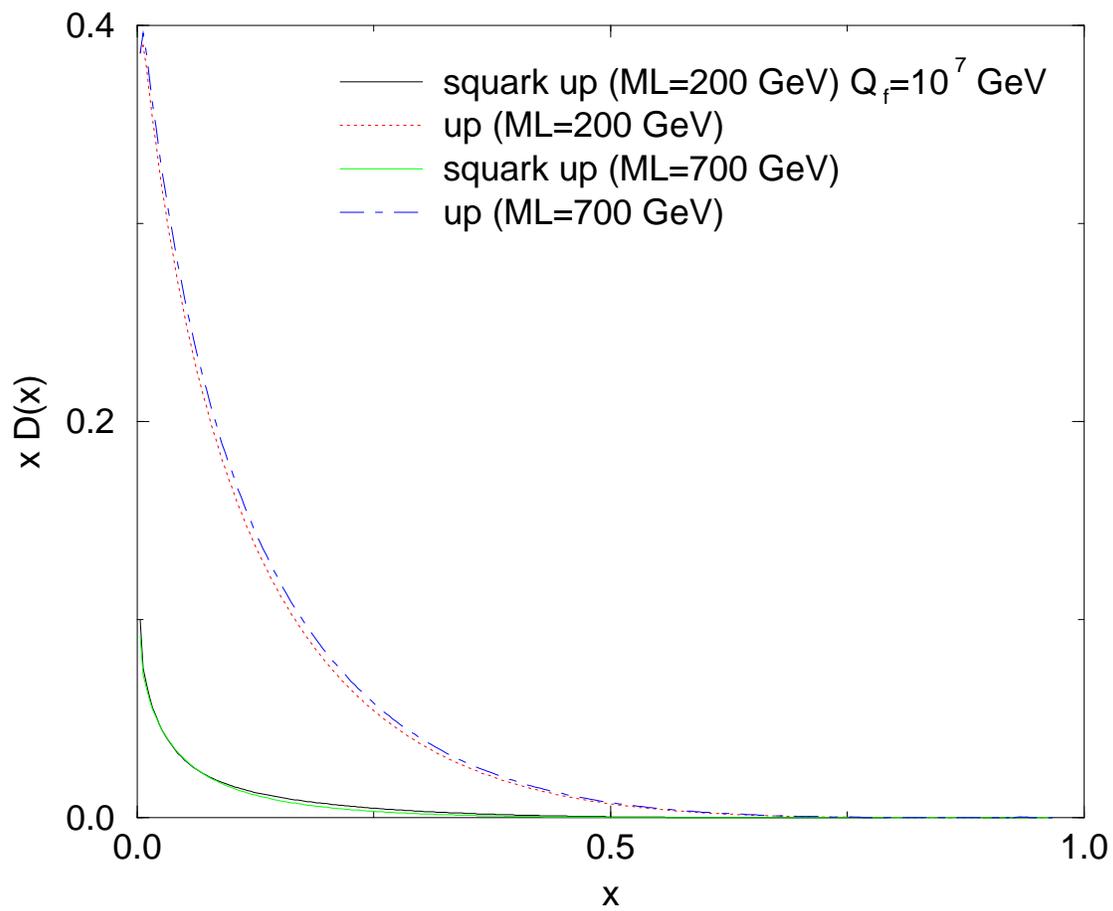}}
\caption{The quark and squark (up) fragmentation functions 
for a varying $m_\lambda$ and a fixed large final scale $Q_f=10^7$ GeV.}
\end{figure}

\begin{figure}
\centerline{\includegraphics[angle=0,width=1.1\textwidth]{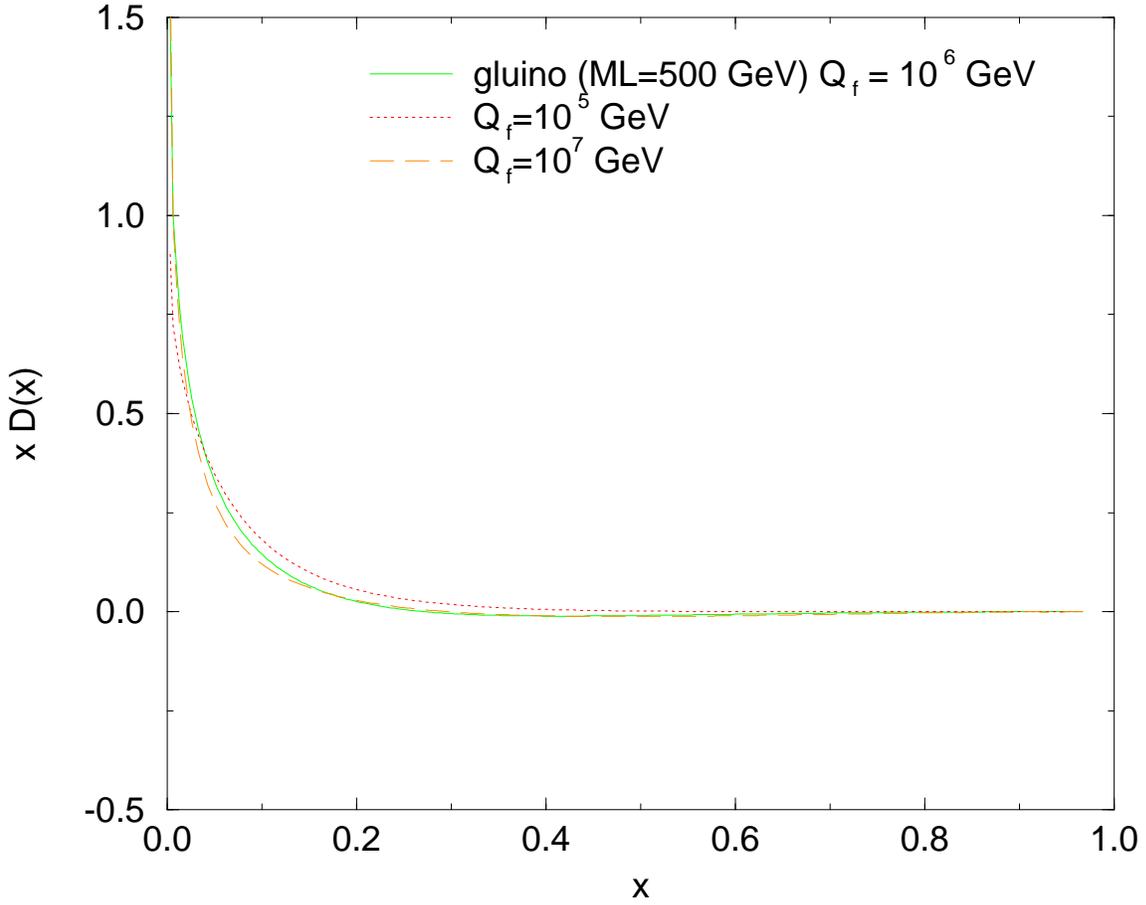}}
\caption{The gluino fragmentation function $x D_{u}^{p,\bar{p}}(x,Q^2)$ 
evolved in the region $500-Q_f$ GeV for different values of $Q_f$ 
The susy fragmentation scale is chosen to be $500$ GeV.}
\end{figure}

\begin{figure}
\centerline{\includegraphics[angle=0,width=1.1\textwidth]{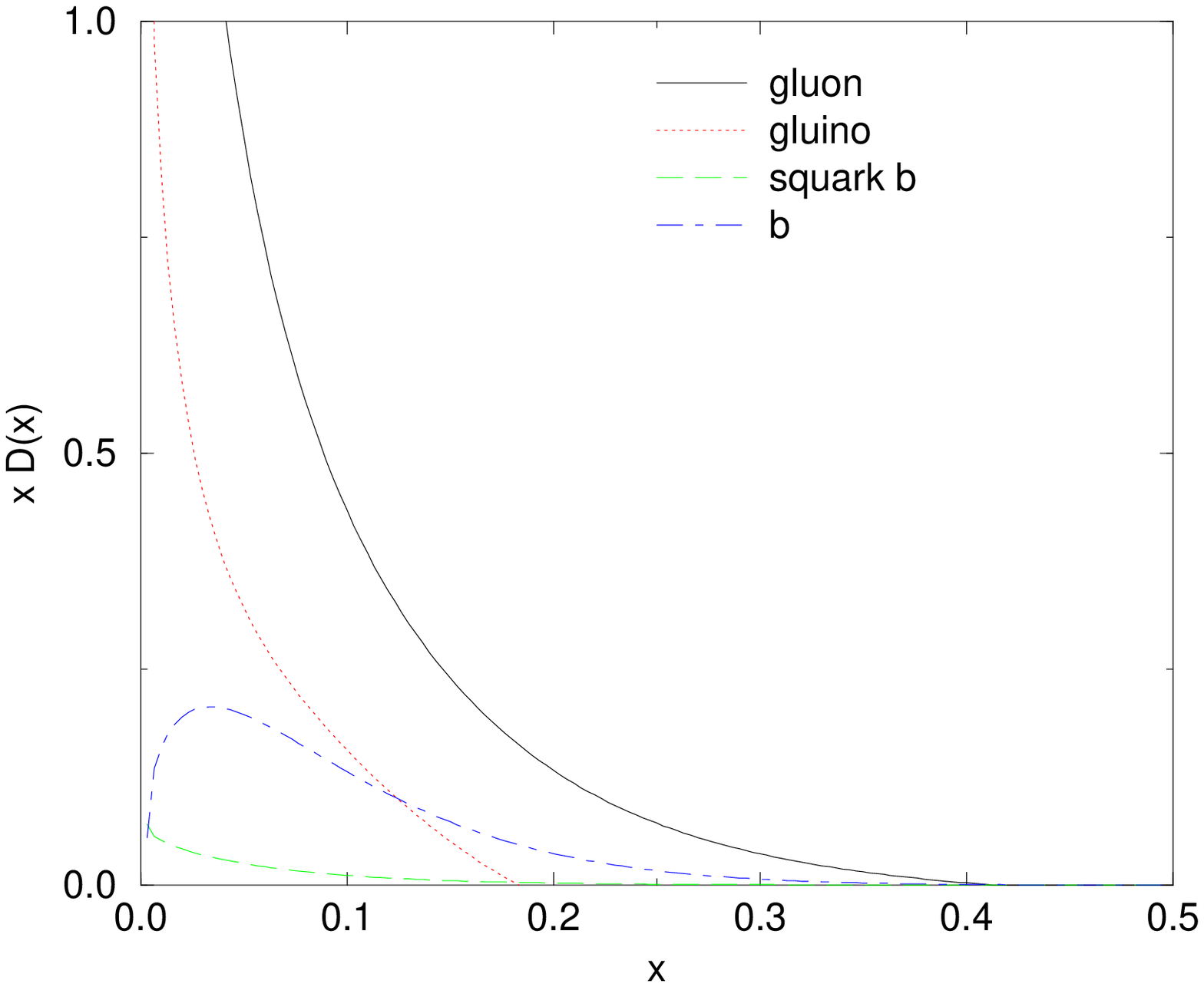}}
\caption{The quark and squark (bottom) fragmentation functions 
together with those of gluon and gluino for $m_\lambda=500$ GeV and a fixed large final scale $Q_f=10^7$ GeV.}
\end{figure}


\begin{thebibliography}{99}
\bibitem{gcurge} J. Ellis, S. Kelley and D.V. Nanopoulos, 
\PLB{260}{1991}{131};\\
P. Langacker and M. Luo, \PRD{44}{1991}{817};\\
U. Amaldi, W. de Boer and H. F\"ustenau, 
\PLB{260}{1991}{447}.
\bibitem{ccf} C. Corian\`{o}, S. Chang and A.E. Faraggi, 
\PLB{397}{1997}{76};
\NPB{477}{1996}{65};\\
K. Benakli, J. Ellis and D.V. Nanopoulos, \PRD{59}{1999}{047301};
C. Corian\`{o}, A.E. Faraggi and M. Pl\"umacher, OUTP-01-35P,
to appear.
\bibitem{Subir1} For review and references see {\it e.g.}: \\
		S. Sarkar, Plenary Talk at COSMO 99, hep-ph/0005256.
\bibitem{CC1} C. Corian\`{o}
{  Supersymmetric Scaling Violations. 1. An algorithm to solve the supersymmetric DGLAP evolution}, { hep-ph/0009227 }
\bibitem{CC2} C. Corian\`{o},{  Supersymmetric Scaling Violations 2. 
The general supersymmetric evolution },{ hep-ph/0102164} 
\bibitem{CC3} C. Corian\`{o}, {  Distributions of $N=1$ QCD}, { hep-ph/0101352}
\bibitem{KR} C. Kounnas and D.A. Ross, Nucl. Phys. {B214}:317, (1983).
\bibitem{Antoniadis} I. Antoniadis, C. Kounnas and R. Lacaze, Nucl. Phys. { B211}:216, 
(1983).
\bibitem{blumlein}J. Blumlein, V. Ravindran and W.L. van Neerven 
Nucl.Phys.B586:349-381, (2000). 
\bibitem{GZK} K. Greisen, Phys. Rev. Lett. 16:748, (1966); G.T Zatsepin and V.A. Kuzmin, 
Sov. Phys. JETP Lett. 4:78 (1966).
\bibitem{kkp} B.A. Kniehl, G. Kramer, B. P\"otter,
Nucl.\ Phys.\ B 582:514, (2000). 
\end{thebibliography}
\end{document}